\documentclass{article}
\usepackage{arxiv}
\usepackage{hyperref}
\hypersetup{
    colorlinks = true,
    urlcolor = {blue},
    citecolor= {black}
}
\usepackage[T1]{fontenc}
\usepackage{amsmath}
\usepackage{cleveref}
\usepackage{graphicx}
\usepackage{placeins}
\usepackage{float}
\usepackage{footnote}
\usepackage{titlesec}
\usepackage{tcolorbox}
\usepackage{CJKutf8}
\usepackage{verbatim}
\usepackage{pdflscape}
\usepackage{colortbl,hhline}
\usepackage{ebgaramond}
\usepackage{hyperref} 
\usepackage{xcolor}   

\hypersetup{
    citecolor=blue,        
}

\usepackage{natbib}

\newcommand{\versiondate}{Jan 23th, 2025}
\newcommand{\versiondatetext}{Previous Version: Aug 28 , Oct 24, and Nov 13th 2024}

\title{Take Caution in Using LLMs as Human Surrogates: \\Scylla Ex Machina\thanks{We thank seminar participants at the BU, Wharton (Sep 2024), USC, UC Irvine, and Meta. All errors are the author’s own.}}

\author{
Yuan Gao \\
  Questrom School of Business \\
  Information Systems Department\\
  Boston University\\
  Boston, MA 02215 \\
  \texttt{yuangg@bu.edu} \\
   \And
 Dokyun Lee \\
  Questrom School of Business \\
  Information Systems Department and\\
  Computing \& Data Sciences \\
  Boston University\\
  Boston, MA 02215 \\
  \texttt{dokyun@bu.edu} \\
  \And
  \quad Gordon Burtch \\
  \quad Questrom School of Business \\
  \quad Information Systems Department\\
  \quad Boston University\\
  \quad Boston, MA 02215 \\
  \quad \texttt{gburtch@bu.edu} \\
    \And
 Sina Fazelpour \\
  Department of Philosophy and \\ Khoury College of Computer Sciences \\
  Northeastern University\\
  Boston, MA 02115 \\
  \texttt{s.fazel-pour@northeastern.edu} \\
}

\begin{document} 

\maketitle
\begin{abstract}
Recent studies suggest large language models (LLMs) can exhibit human-like reasoning, aligning with human behavior in economic experiments, surveys, and political discourse. This has led many to propose that LLMs can be used as surrogates or simulations for humans in social science research. However, LLMs differ fundamentally from humans, relying on probabilistic patterns, absent the embodied experiences or survival objectives that shape human cognition. We assess the reasoning depth of LLMs using the 11-20 money request game. Nearly all advanced approaches fail to replicate human behavior distributions across many models. Causes of failure are diverse and unpredictable, relating to input language, roles, and safeguarding. These results advise caution when using LLMs to study human behavior or as surrogates or simulations.
\end{abstract}

\begin{flushright}
    \textit{‘She has twelve misshapen feet, and six necks of the most prodigious length;\\ and at the end of each neck she has a frightful head with three rows of teeth in each’}\\
    --- Homer, \textit{Odyssey} (Describing Scylla)
\end{flushright}

\section*{Introduction}

Recent studies report that Large Language Models (LLMs) can exhibit human-like cognitive abilities. These studies demonstrate that LLMs show behaviors that align closely with those of human subjects in seminal experiments from behavioral economics, and responses comparable to those of humans in perceptual tasks and standard theory of mind (ToM) evaluations \citep{r1,r120,r121}. Early work in this area \citep{r68} suggests that LLMs can mirror human preferences and biases in decision-making and behavior more generally across a variety of domains, including political science \citep{r78}, marketing \citep{r70}, and psychology \citep{r67, r69, r74}.

On this basis, studies claim that LLMs may serve as useful surrogates or simulation testbeds for human subjects in social science research \citep{r122,tranchero2024theorizing}, an idea that has also gained traction in industry, e.g., for market research \citep{r70}.\footnote{https://www.kantar.com/north-america/inspiration/analytics/what-large-language-models-could-mean-for-market-research} For example, the startup Synthetic Users\footnote{https://www.syntheticusers.com/} employs OpenAI models to simulate consumer reactions to new products. The benefits of using LLMs as human surrogates have been argued from several perspectives, including cost-effectiveness, scalability \citep{arora2024express}, and the idea that LLMs can provide synthetic data in cases where it has traditionally been scarce \citep{r1}. 

The growing enthusiasm for using LLMs as human surrogates in research raises two urgent questions for the scientific community. First, do findings demonstrating LLMs' capability as surrogates for human participants generalize to other, less explored, or newer research contexts? Second, to what extent are LLMs' behaviors sensitive to design and implementation choices? To the extent the behavior of LLMs does depend on these choices, we argue they require careful documentation and evaluation by researchers and reviewers invested in ideals of replicable and responsible science \citep{messeri2024artificial, burnell2023rethink, burden2024evaluating}. This work investigates these issues. 

We begin with an empirical case study demonstrating the instability of several LLMs in a simple experiment, wherein we show that the models' responses systematically and significantly diverge from those of human participants. Our experiment utilizes the 11-20 money request game \citep{r48}, a simple economic game designed to evaluate participants' depth of strategic reasoning.\footnote{We chose this game because i) it is more recent than classic games (e.g., the beauty contest game) and thus is less likely to have been memorized by the LLMs under investigation, and ii) the game is exceedingly simple, such that a failure to exhibit human-like behavior here would be the cause for serious concern about LLM behavior in more complicated scenarios.} We begin by exploring variation in responses to the experiment instructions across eight popular LLMs (GPT-4, GPT-3.5, Claude3-Opus, Claude3-Sonnet, Llama3-70b, Llama3-8b, Llama2-13b, and Llama2-7b). We collect responses from 1,000 clean sessions for each LLM and compare the distribution of those responses to the distribution of previously reported human participant responses and Nash Equilibrium predictions. We demonstrate that LLMs' behavior differs markedly from that of human participants. Moreover, we show significant variation in response distributions \textit{across} LLMs. Notably, the most recent and largest models do not necessarily exhibit more human-like reasoning, a finding that runs counter to the notion that more human-like responses may be induced or enabled in LLMs through scaling \citep{r118}. And, comparing LLM responses with those of humans across game variants specifically designed to elicit certain reasoning modes, we find that LLMs do not display the same level of sensitivity to variations in game design as humans.


Next, we extend the experiment by applying three advanced techniques that have been proposed to enhance LLM performance in reasoning. Broadly, we demonstrate that LLMs exhibit unstable behavior that differs from human behavior to a statistically significant degree, regardless of the approach used. Specifically, our analysis reveals a few key findings. First, all out-of-distribution (OOD) approaches\footnote{OOD approaches refer to the evaluation of LLM behavior absent the explicit provision of human behavioral examples via prompts or fine-tuning, and ensuring that the model has not memorized prior patterns of human response to the particular task in its parameters.} to eliciting human-like behaviors generally fail, regardless of the LLM used. Second, providing explicit human behavioral examples as part of prompting or via Retrieval-Augmented Generation (RAG) or fine-tuning, makes LLM outputs slightly more similar to human distributions. However, surprisingly, none of these approaches, with any LLM models, is fully capable of inducing behavior that replicates human-like distributions, with the sole exception of a fine-tuned GPT-4o. Even so, we believe fine-tuned GPT-4o only mimics the data we used on human results as documented in the paper that first introduced the 11-20 money request game, and is limited to the specific patterns and contexts it has encountered. This limitation is further reinforced through experiments on Centaur \citep{binz2024centaurfoundationmodelhuman}, a newly released fine-tuned model based on Llama 3.1, which was trained on extensive human data to explicitly emulate human behavior. Despite this, Centaur still fails to replicate human behavioral patterns in our game.

We further show that LLMs' responses vary significantly based on the provision of in-line examples during prompting, as well as with changes in prompt framing, even when the game instructions remain unchanged. For example, instructions to respond `rationally' versus `like a human' lead to different response distributions, as does changing the language (English, Chinese, Spanish, and German). This underscores the instability of LLM outputs and the issue of prompt brittleness.

Having established the behavioral instability of LLMs in a simple setting, and their divergence from human-like behavior, we investigate the underlying causes, querying LLMs to explain their choices. We find discrepancies between LLMs' stated rationale and their actual behavior, and we observe that some advanced (larger) models even appear to misunderstand the game instructions, despite the game's simplicity. Further, we consider the degree to which models' responses may depend on their prior memorization of the experiment and past findings. Whereas the various LLMs appear incapable of providing the instructions associated with the experimental game we consider here (an indicator that the game has not been memorized by these models), each is generally \textit{perfect} at recalling the instructions associated with classic economic games that evaluate the same level-k reasoning abilities as our game. This observation, along with their divergence from human samples' distribution and failure in OOD scenarios, suggests the potential role of memorization as a driver of LLMs' previously documented ability to mimic human-like behavior in certain experiments.

We contextualize our findings within the broader discussion of LLM failure modes, which may arise from the inherent differences between LLM objective functions---focused on memorizing and predicting the statistical regularities of online textual data---and human objective functions---centered on survival and reproduction---as discussed in the literature on the philosophy of AI \citep{r155,r159}. We provide an overview of recent empirical research on LLMs, highlighting inherent flaws, such as performance inconsistency, reliance on memorization, and prompt brittleness, all of which significantly limit the feasibility of using LLMs in studying human behavior or as human surrogates.

Lastly, drawing on these experiments and emerging explanations, we turn to our second question, to examine the different choices and assumptions that are inherently involved when LLMs are deployed as human surrogates, from experiment design to interpretation, and thus require careful documentation. We identify a series of questions and criteria that should be borne in mind by researchers, reviewers, and other stakeholders, to evaluate the stability of LLM behavior across models, conditions, and replications. 


\section*{LLMs Fail in a Simple Experiment}

\subsection*{11-20 Money Request Game and Level-k Reasoning}

Level-k thinking is a theoretical framework used in game theory and behavioral economics to model strategic decision-making in interactive settings \citep{r117}. It suggests that individuals operate at varying reasoning levels and base their strategies on iterative predictions about others' reasoning depth. This framework is particularly effective for analyzing real-world strategic interactions where perfect rationality and complete information are unrealistic. Empirical evidence indicates that Level-k models often accurately reflect human behavior in experimental settings \citep{r116}. Under this framework, prior studies have demonstrated that LLMs exhibit a reasoning depth comparable or superior to that of humans in the classic beauty contest game \citep{r125,r126,r129}. This framework relates closely to the Theory of Mind (ToM), which numerous studies have also drawn upon in the study of LLMs' reasoning capabilities, many concluding that LLMs are more capable of inferring others' mental states than humans \citep{r121,r162}.

This paper evaluates LLMs' reasoning depth employing the 11-20 Money Request Game, an experimental game designed to test level-k reasoning. This game has several desirable characteristics for our purposes. First, the 11-20 money request game is less well-known and thus less likely to have been memorized by popular LLMs than more established, more popular experiments such as the Beauty Contest Game. This aspect allows us to i) explore LLMs' relative performance in a setting where memorization is less likely to influence LLM performance, and ii) examine how intentionally inducing LLMs' memorization of the game impacts their performance. Second, the 11-20 money request game is a very simple experiment, reducing any possible role of instruction (mis)comprehension or (mis)understanding as a driver of performance. Because of its simplicity, this game naturally evaluates level-k thinking without inducing other decision rules or mental processes that may confound results. The game instructions are as follows:

\textit{You and another player are playing a game in which each player requests an amount of money. The amount must be (an integer) between 11 and 20 shekels. Each player will receive the amount he requests. A player will receive an additional 20 shekels if he asks for exactly one shekel less than the other. What amount of money would you request?}
\par

The number 20 is a natural anchor for the iterative reasoning process in this game. It is the intuitive choice within the 11 to 20 shekel range, as 20 is the highest and most visible number. Choosing 20 also appeals to players who prefer to avoid risk or strategic thought, as it offers the highest guaranteed amount. The evaluation of level-k thinking in this game is straightforward: selecting 19 in response to 20 (level-1), 18 in response to 19 (level-2), and so forth, down to 11, reflects the depth of a respondent's strategic reasoning. Each step in this reasoning chain builds on the expected actions of the opponent, reflecting increasingly complex strategic layers. 

We employ eight popular LLMs
(GPT-4, GPT-3.5, Claude3-Opus, Claude3-Sonnet, Llama3-70b, Llama3-8b, Llama2-13b, and Llama2-7b) and repeat the experiment procedure 1,000 times with each,\footnote{We control the model's temperature at 0.5, which keeps a balance between output coherence and diversity; approximately the same results are observed with a temperature of 1.} instantiating separate (independent) sessions with each iteration. We collected the data between February and April 2024.\footnote{Further details about the models and settings can be found in \textit{Appendix}}

\subsection*{Results}
\begin{figure}
\centering
\includegraphics[width=0.8\linewidth]{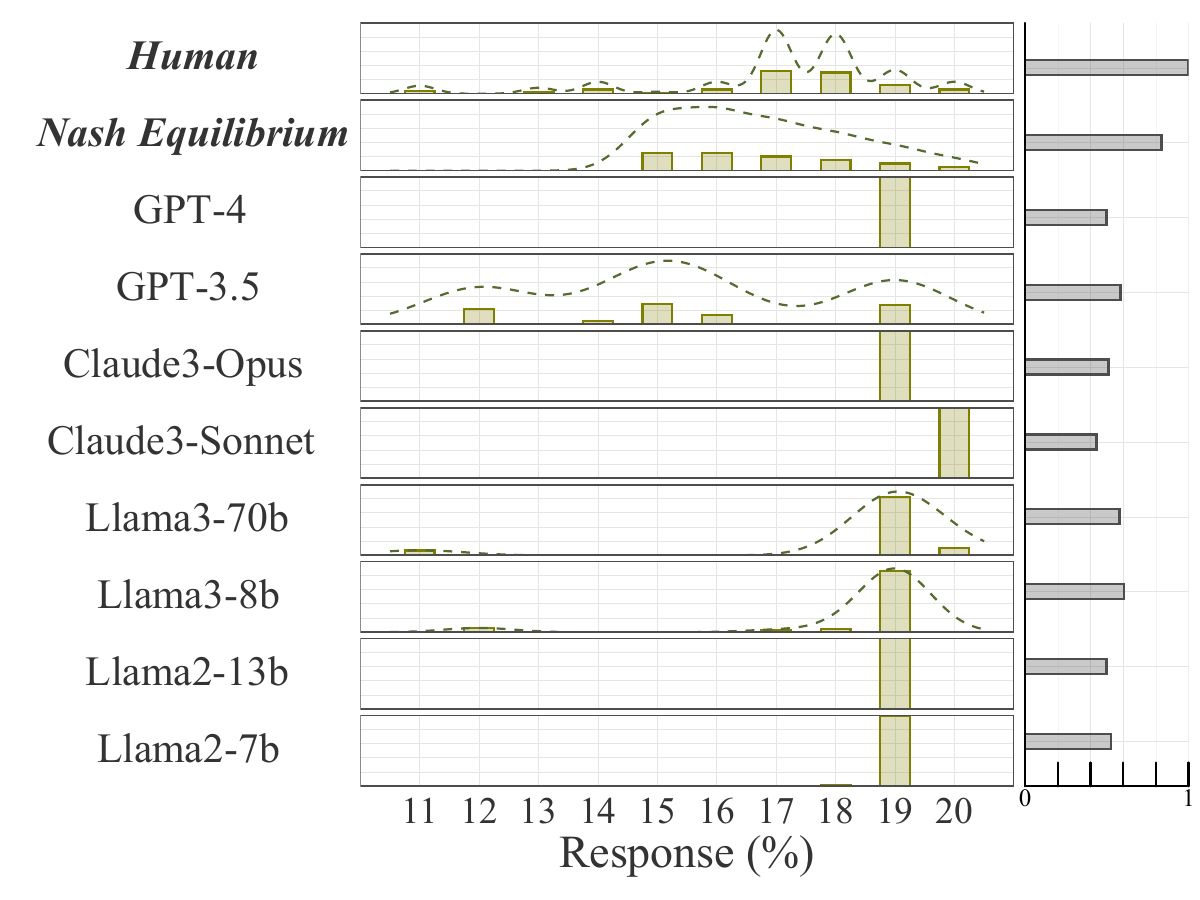}
\caption{\textbf{11-20 Money Request Game.} The bar chart on the right shows the similarity between the distribution of different subjects and human subjects, measured by Jensen-Shannon divergence scores. Density plots are omitted for subjects with over 98\% of the data concentrated in a single choice to avoid potential misinterpretation.}
\label{org}
\end{figure}

We compare the response distributions of LLMs with previously reported results obtained from human subjects and Nash Equilibrium predictions \citep{r48}.\footnote{Here, the human results are those associated with 108 human players who participated in the original experiment as documented in the manuscript that introduced the 11-20 money request game \citep{r48}. The Nash Equilibrium predictions come from the same source.} We arrive at several notable findings, as shown in Fig.~\ref{org}.

First, although LLMs typically excel in classic games that seek to evaluate the same form of strategic thinking, we find that LLMs' performance differs here. In the beauty contest game, LLMs have been found to exhibit a reasoning depth comparable or superior to that of humans \citep{r125,r126,r129}. By contrast, here they consistently demonstrate a lower level of reasoning than humans. Our findings reveal that all advanced LLMs, except for GPT-3.5, tend to select values of 20 or 19, corresponding to level-0 and level-1 reasoning, respectively. This level of strategic reasoning is two levels below that of typical human participants, i.e., level-3 reasoning, or the selection of the number 17 \citep{r48}. 

Second, the broader distribution of LLMs' responses over the 1,000 sessions does not reflect the distribution of human-like responses. Comparing the human and LLM-generated distributions, we observe that all LLMs' response distributions generally diverge from those of humans to a statistically significant degree (permutation test based on Jensen-Shannon divergence, \textit{p} $<$ 0.001, \textit{Appendix Table \ref{js_divergence}}).\footnote{More details about the permutation test can be found in the supplementary Materials.}

Third, the response patterns across LLMs vary greatly, and larger models are not necessarily more human-like than smaller models. For example, GPT-4, despite its broad acceptance as an advancement over GPT-3.5, displayed more limited reasoning depth. Further, despite its ostensibly deeper levels of reasoning, GPT-3.5 exhibited notably diverse responses across sessions, spanning all reasoning levels and exhibiting greater variation than human participants. Our results indicate that increased scale and sophistication in LLMs do not necessarily translate to more human-like behavior. The issue of model selection may become increasingly complex due to various factors, including ongoing model iterations and implicit updates within versions. For instance, recent research indicates that larger, instruction-tuned language models are less reliable than smaller models \citep{zhou2024larger}.

Fourth, and last, we observe from \textit{Appendix Fig. \ref{diff_version}} that some of the LLMs are not responsive to changes in task design that are specifically formulated to trigger changes in modes of reasoning, while others are responsive, yet in ways that make them less human-like. The 11-20 Money Request Game has two variants designed to explore human decision-making biases that limit deeper reasoning. The first variant of the game that we consider is the cycle variant, which enhances the significance of the level-0 strategy by allowing the choice of 20 shekels to still yield the 20 shekel bonus when the other player selects 11 shekels. This modification makes 20 shekels an even more appealing and justifiable choice, as a starting point for iterative reasoning. In this variant, humans tend to choose 19 more often but with reduced overall reasoning depth. Claude-3-Sonnet shows a similar shift to 19 but with increased overall reasoning depth, unlike humans. GPT-3.5 shifts unexpectedly toward 12, indicating level-8 reasoning. The most advanced models, GPT-4 and Claude-3-Opus show no change in their response distributions.

The second variant of the game is the costless variant, which minimizes the impact of risk preferences by ensuring that if participants choose any number between 11 and 19, their losses remain fixed regardless of the opponent's choices. In this version, the distribution of human participants' responses does not change significantly relative to the basic version of the game. By contrast, GPT-3.5 and Claude-3-Sonnet shift their responses markedly in the costless variant. Once again, with the most advanced closed-source models, GPT-4 and Claude-3-Opus, we see no changes in their response distributions relative to the basic version of the game.

\section*{LLMs Rarely Mimic Human Behavior, Even with Examples}
In the previous sections, we highlight key differences between human behavior and that of LLMs in `out-of-the-box' deployment, most notably demonstrating that LLMs have significantly lower reasoning depth than humans. In this section, we explore the possibility that more advanced techniques could be used to enhance LLM reasoning to better align it with that of humans. Specifically, we apply three commonly used methods to improve LLM performance in reasoning tasks: prompt engineering \citep{r147}, fine-tuning \citep{r153}, and retrieval-augmented generation (RAG) \citep{r152}.

\begin{figure*}
\centering
\includegraphics[width=0.9\textwidth]{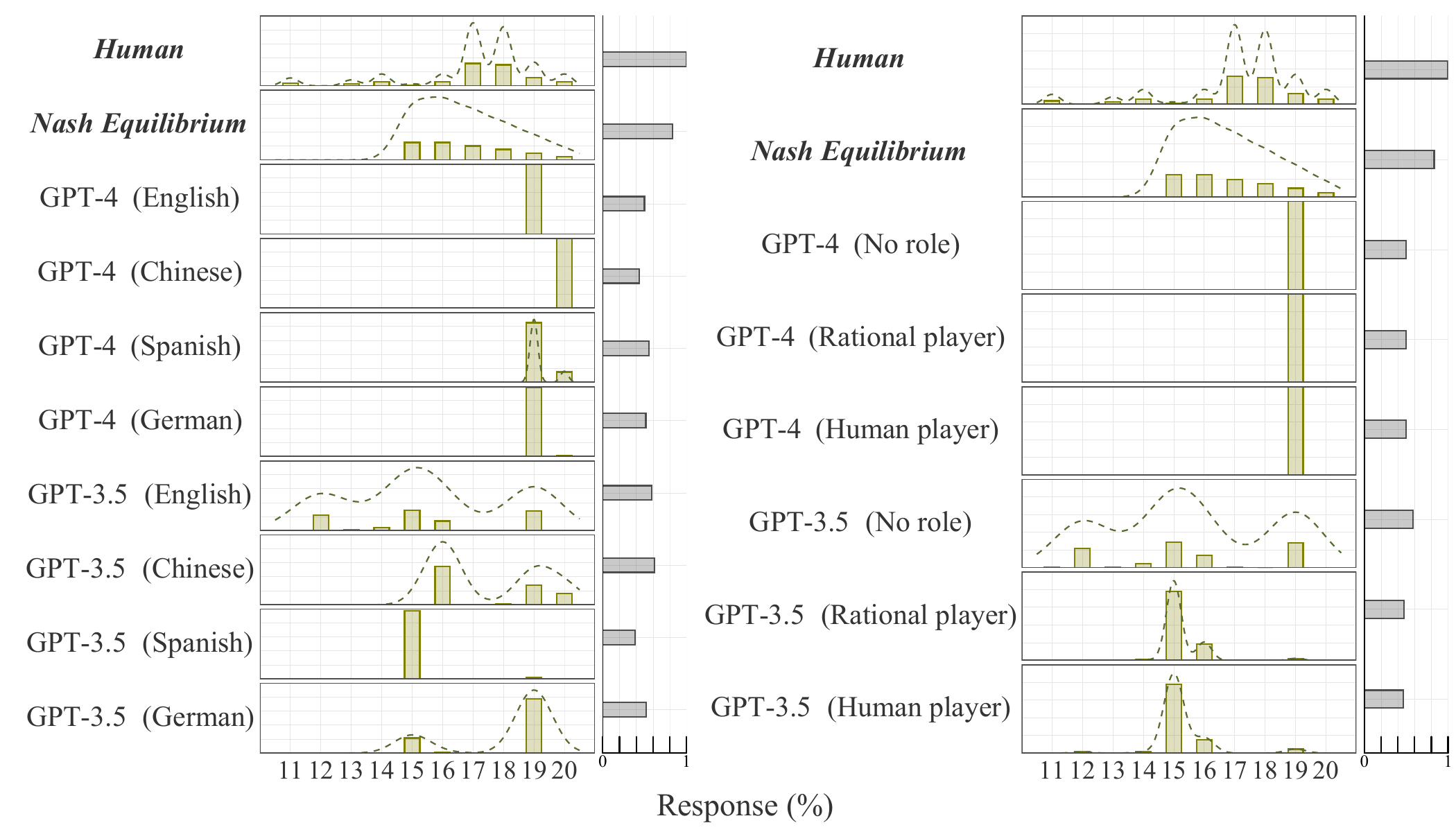}
\caption{\textbf{Prompt Brittleness: Roles and Languages.} The bar chart on the right shows the similarity between the distribution of different subjects and human subjects, measured by Jensen-Shannon divergence scores. Density plots are omitted for subjects with over 98\% of the data concentrated in a single choice to avoid potential misinterpretation.}
\label{brittle_all}
\end{figure*}

\subsection*{Zero-shot Prompts}
Zero-shot prompting is an approach that seeks to improve LLM performance by carefully designing the input prompts without providing any additional context and examples. The model carries out `inferences' based strictly on its internal parameters, as learned during initial training. Well-crafted prompts are regarded as effective tools to guide LLMs in understanding tasks and reasoning appropriately. For instance, one popular prompt engineering technique we will use, Chain-of-Thought (CoT) \citep{r144}, instructs LLMs to break down tasks into smaller steps and think through them sequentially to arrive at a final answer \citep{r144}. Beyond this broader idea, specific research has also found that providing LLMs with an instruction to `take a deep breath and work on this problem step-by-step' \citep{r146}, or providing LLMs with emotional stimuli like `This is very important to my career' \citep{r145} can enhance task performance. We consider all three of these zero-shot strategies here. 

\begin{figure}[htbp]
\centering
\includegraphics[width=1\textwidth]{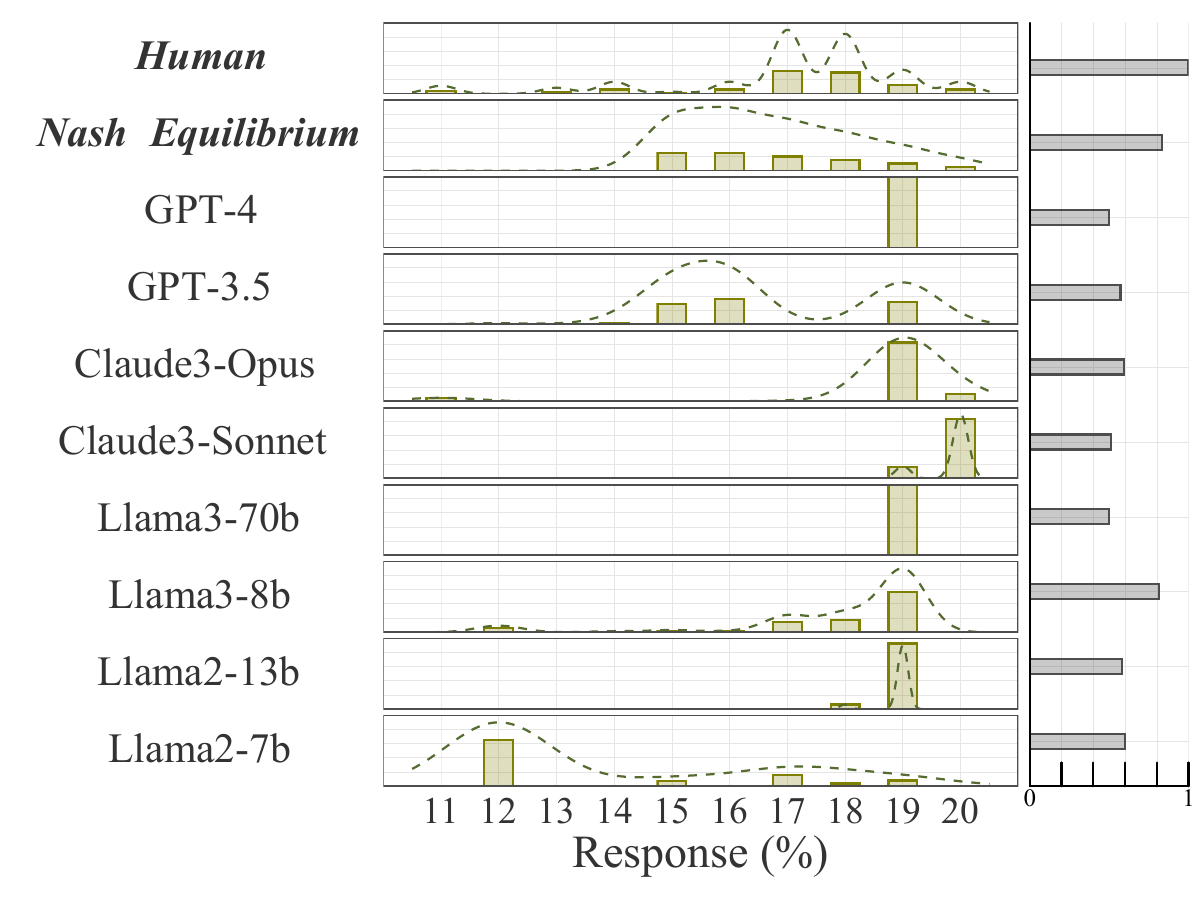}
\centering
\caption{\textbf{Advanced Prompting: Optimization Based.} The bar chart on the right shows the similarity between the distribution of different subjects and human subjects, measured by Jensen-Shannon divergence scores. Density plots are omitted for subjects with over 98\% of the data concentrated in a single choice to avoid potential misinterpretation.}
\label{zs-optimization}
\end{figure}

Our results are presented in \textit{Appendix Fig.~\ref{zs-cot}, \ref{zs-emotion}, and~\ref{zs-optimization}}. We find that, in most cases, these strategies do not induce more human-like responses from LLMs. For the largest, most advanced models, these techniques have very little effect. Among some of the smaller models, such as llama3-8b and llama2-7b, we do see changes in models' output distributions. However, those changes do not amount to more human-like behavior. For instance, under the influence of optimization-based advanced prompts, that is, `take a deep breath' prompt, llama2-7b's output distribution becomes more concentrated around numbers representing very deep levels of reasoning, levels rarely observed in human samples. 

Additionally, previous research presents some conventions in zero-shot prompt design that lack careful documentation. For example, researchers from different countries use various languages as prompts, and some include additional `statements' in the prompts like `supposing you are a human/rational player'. Considering that LLMs are suggestible, and their behavior significantly shifts based on question phrasing \citep{r92}, we explore whether presenting games in different formats, without altering the rules, leads to behavioral changes in LLMs. We test this by assigning roles and using various languages in prompts. LLMs are designated as either rational or human players, and we examine their responses to prompts in English, Chinese, Spanish, and German.

Our findings displayed in Fig.~\ref{brittle_all} are notable: language and role significantly influenced model behavior, despite the game rules being unchanged. Specifically, GPT-3.5, when prompted in Chinese, focuses more on intermediate reasoning depths than English prompts, while GPT-4 shows shallower reasoning in Chinese than in English. Additionally, assigning different roles causes substantial behavior changes in GPT-3.5, reducing output diversity and focusing on the reasoning depth rarely observed in human samples. This raises the concern that researchers may opt for certain LLM roles or input languages without being fully aware of the consequences for LLMs' behavior, mistakenly believing they are observing LLMs' `true' behavioral patterns, when in fact the observations are highly unstable and dependent on seemingly innocuous choices.

We further test whether instantiating LLMs with demographic personas within prompts, as suggested by prior studies \citep{r68}, enables LLMs to generate responses that align with a human sample. \footnote{The demographic information we consider includes gender, age, race, employment status, marital status, income, education, and religion.} We demonstrate this by collecting 143 additional responses from US participants on Prolific, paired with their demographic data. We then include this information when obtaining responses from the LLMs. As shown in \textit{Appendix Fig.~\ref{prolific}}, including demographic personas does not induce greater diversity in LLM responses, and LLMs' response distributions do not become more human-like.

\subsection*{Few-shot Prompts}
We next turn our attention to few-shot prompting techniques, i.e., providing a few examples for LLMs to learn from `on the fly', without updating model parameters \citep{r147}. Notably, one recent study has shown that few-shot learning can synthesize more human-like responses for market research \citep{arora2024express}. We explore this possibility here, specifically employing CoT prompting. We provide the LLMs with three exemplary answers that include response values along with step-by-step reasoning that would rationalize those choices. To avoid selection biases, our samples include both commonly selected values from human samples, i.e., 18, 19, and 20, and less common choices, i.e., 11, 12, and 13. These choices correspond to the highest and lowest levels of reasoning depth.

\begin{figure*}
\centering
\includegraphics[width=0.9\textwidth]{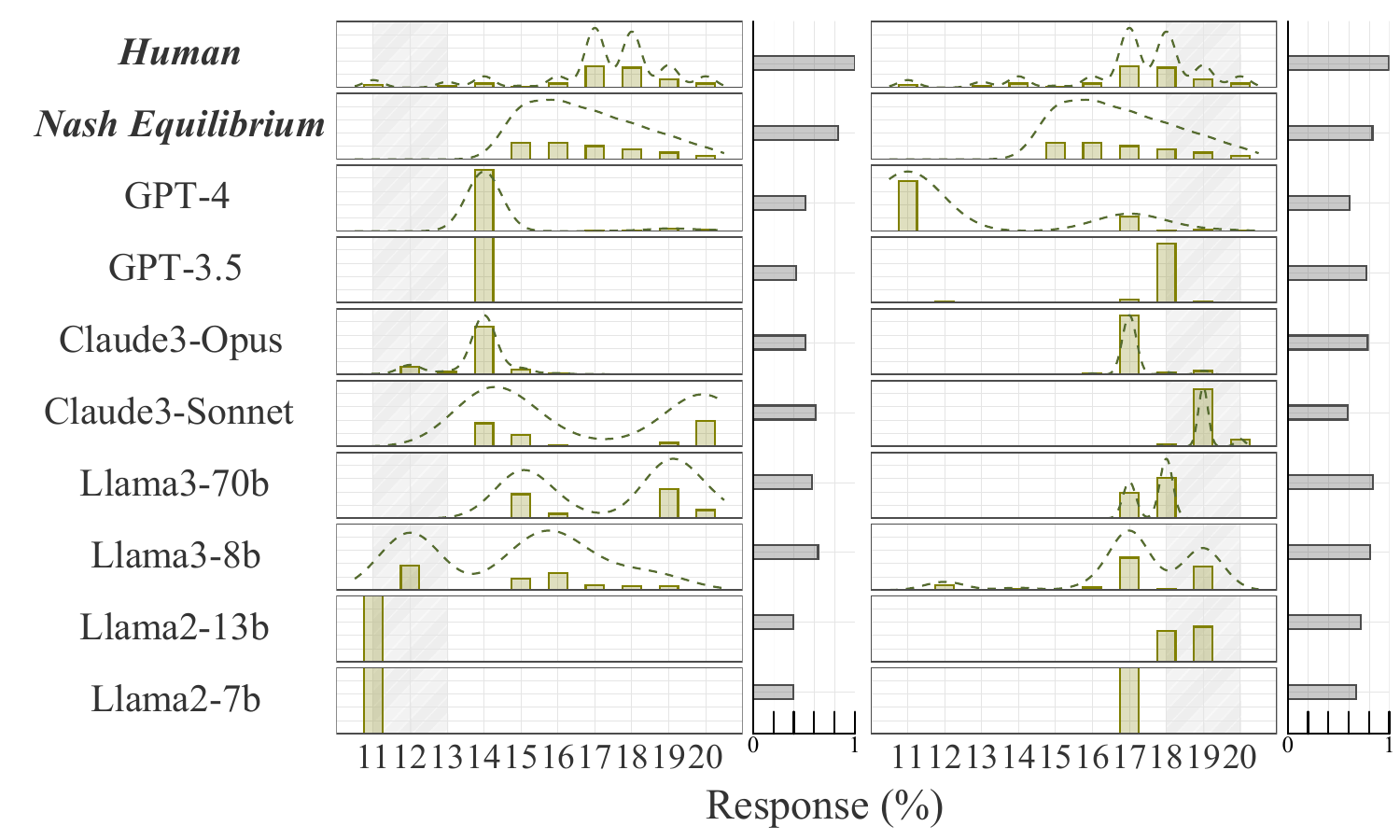}
\caption{\textbf{Few-shot CoT.} The shaded gray area represents the sample range we provided. The bar chart on the right shows the similarity between the distribution of different subjects and human subjects, measured by Jensen-Shannon divergence scores. Density plots are omitted for subjects with over 98\% of the data concentrated in a single choice to avoid potential misinterpretation.}\label{cot_all}
\end{figure*}

As shown in Fig.~\ref{cot_all}, LLMs are highly sensitive to examples that are provided. In general, the models appear to merely copy the examples; when presented with example responses that reflect higher or lower levels of reasoning, the output distributions of all LLMs typically cluster around the examples.

\subsection*{Retrieval Augmented Generation (RAG)}

RAG operates by providing LLMs with access to domain-specific external knowledge for inference. Specifically, RAG refers to models' retrieval and incorporation of relevant content from provided external documents when producing a response. We embed the original manuscript that introduced the 11-20 Money Request Game \citep{r48}, making it available as external knowledge to both GPT-4 and GPT-3.5. 

Figure~\ref{posttrain} shows that, with the help of RAG, LLMs' behavior becomes more aligned with human behavior. However, as indicated in \textit{Appendix Table \ref{js_divergence}}, the resulting distribution remains statistically significantly different from that of humans.

\subsection*{Fine-tuning}
Fine-tuning aims to enhance model performance via a logic similar to few-shot prompting, i.e., by inducing the model to learn from examples. However, during fine-tuning, the model's internal parameters are permanently altered, potentially influencing all subsequent inferences, regardless of prompt structure. Fine-tuning usually requires a substantial amount of data and may not always successfully induce LLM output changes.

To achieve this, we create a dataset, which replicates the entire set of choices and reasons provided by prior human subjects as published in the prior manuscript that introduced the 11-20 money request game, to perform fine-tuning.\footnote{More details can be found in the Supplementary Materials.} Compared to the examples we provided in the few-shot CoT prompts, the new dataset offers a broader, more diverse set of examples and incorporates `guessing' as a basis for the choice —an irrational reason—which is found in human samples \citep{r48} and has the potential to bring the LLM closer to the real-world human decision-making process.\footnote{Since the version of GPT-4 we have tested up to this point does not support fine-tuning, we leverage a more recent model for this purpose, namely GPT-4o, which has equivalent performance to GPT-4-Turbo in text processing tasks: https://openai.com/index/hello-gpt-4o/.}

\begin{figure}
\centering
\includegraphics[width=0.8\linewidth]{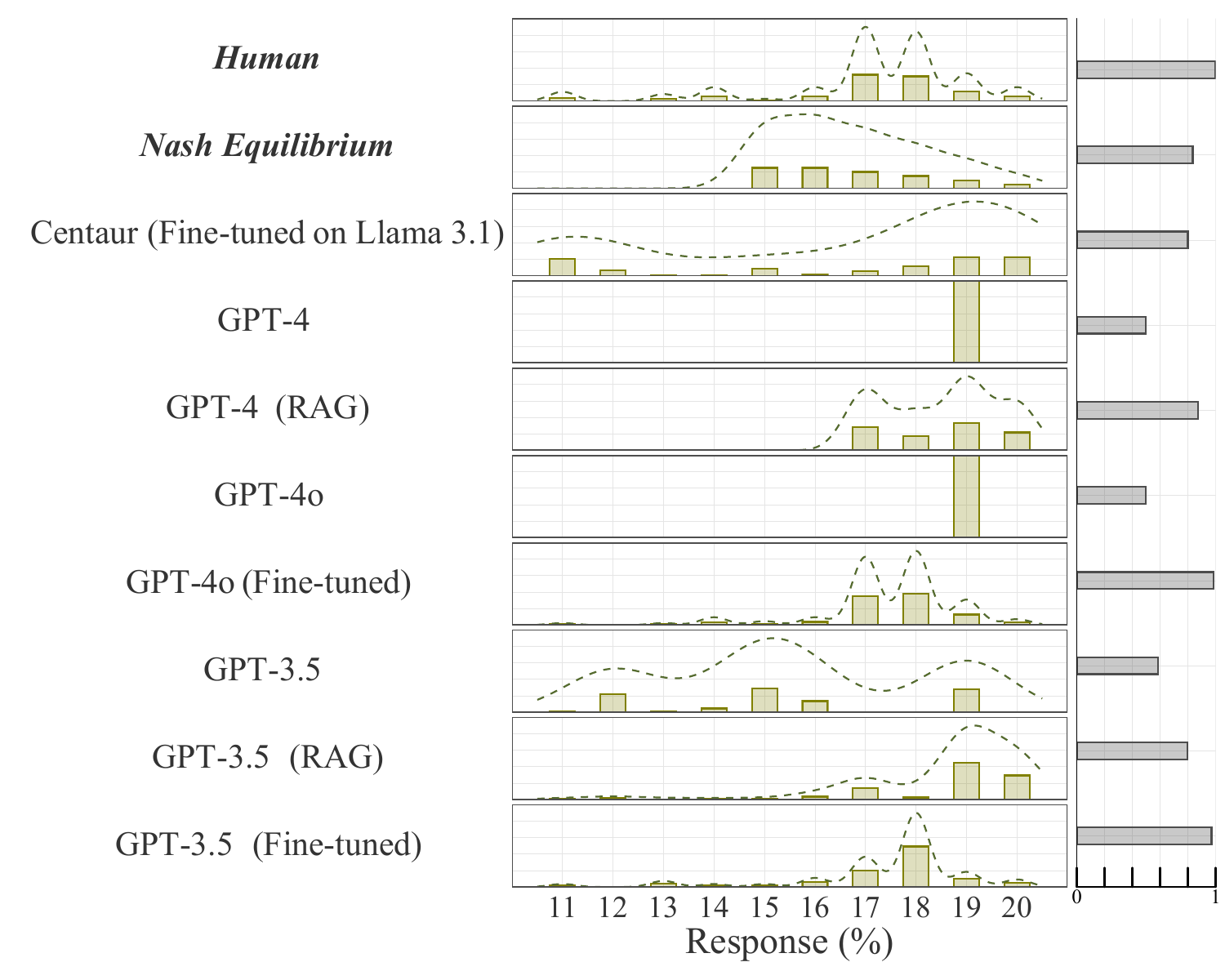}
\caption{\textbf{RAG and Fine-tuning.} The bar chart on the right shows the similarity between the distribution of different subjects and human subjects, measured by Jensen-Shannon divergence scores. Density plots are omitted for subjects with over 98\% of the data concentrated in a single choice to avoid potential misinterpretation.}\label{posttrain}
\end{figure}

Consistent with findings from the few-shot CoT prompting evaluation, as shown in Fig.~\ref{posttrain}, LLMs tend to replicate patterns from the provided samples. However, unlike with few-shot CoT, fine-tuning allows GPT-4o to produce a distribution that could not be statistically significantly rejected under the null hypothesis that it comes from the same distribution as humans, as shown in \textit{Appendix Table \ref{js_divergence}}. It is important to note that these results do not provide any indication of how GPT-4o would generalize to unseen data or novel situations, with or without fine-tuning.

To emphasize these insights, we extend our analysis to a newly released fine-tuned model, Centaur \citep{binz2024centaurfoundationmodelhuman}, designed to align LLM outputs with human-like distributions. Centaur exhibits a diverse response distribution, yet it still deviates from typical human behavior. This suggests that fine-tuned models, despite being trained on extensive human behavioral data, remain challenged in capturing the complexity of human cognition and in generalizing to broader human-centered experiments.

Furthermore, fine-tuning necessitates a substantial amount of diverse, high-quality, and representative data, resulting in a paradoxical situation that undermines the rationale and intention of using LLMs to simulate samples when human data is scarce. 

Overall, providing LLMs with examples appears to lead to a pronounced demand effect \citep{r160}. That is, the examples provided do not appear to `teach' the LLM to reason like a human subject. Instead, the LLMs appear to draw on the examples to discern what type of response the user expects to see, and then the model delivers on those `cues', producing the expected response.

\section*{Understanding LLMs Failure Modes}
Our experiments expose substantial inconsistencies across the LLMs we test, arising from model variations, experimental designs, and prompts. To better understand when and why these inconsistencies occur, we further investigate by interviewing LLMs. Given the limited explainability of LLMs, prior work has found it useful to directly ask these models as a means of elucidating their behavior \citep{r92,r65,r66}. 

In LLMs' self-explanations for their choices shown in \textit{Appendix Table \ref{interview}}, we identify several failure modes. Claude3-Opus selects larger numbers to ensure fairness, believing this allows opponents to gain more. This indicates how RLHF-induced (Reinforcement Learning from Human Feedback) attributes like harmlessness and fairness influence LLM behavior more generally. One of the most advanced models, GPT-4, shows behavioral inconsistencies; it claims to choose 19 because it is loss-averse, yet it does not opt for smaller values in the costless variant of the game as shown in \textit{Appendix Fig. \ref{diff_version}}, wherein losses are fixed.

Additionally, many LLMs fail to grasp the instructions despite the game's simplicity. For example, GPT-3.5 often selects 12, 15, and 16, mistakenly believing these are the highest values that would ensure a profit (the highest value that guarantees a profit is, in fact, 20). Similarly, Claude-3-Sonnet and Claude-3-Opus incorrectly assume that requesting one more shekel, rather than one less, will yield a bonus. These errors are unexpected, given LLMs' superior performance on various benchmarks and classical economic experiments.

These failure modes, combined with evidence that LLMs struggle in our simple game and OOD scenarios, raise questions about whether their human-like behavior reflects true reasoning or mere pattern matching based on prior training data, as seen in mathematical reasoning tasks \citep{mirzadeh2024gsm}. To assess this, we test LLMs' ability to reproduce instructions for the 11-20 request game and the beauty contest game, repeating the request 100 times per model. Upon manual verification of the responses, as shown in \textit{Appendix Fig.~\ref{memo_instruction}}, we find that LLMs achieved 75-100\% accuracy in the beauty contest game, yet near 0\% accuracy in the 11-20 request game.\footnote{The only LLM that exhibited an accuracy greater than 0\% in reproducing instructions for the 11-20 money request game was Claude3-Sonnet. However, even then, the model only achieved an accuracy of 2.9\%.} This suggests that classic economic games may not reliably test LLMs' validity as human surrogates, as their apparent human-like behavior may stem from regurgitated training data rather than genuine reasoning.

Given the complexity and limited interpretability of LLMs, their unpredictable and inconsistent patterns are challenging to address, and systematic evaluations are difficult. However, discussions deeply rooted in the philosophy of AI \citep{r150}, along with recent research on LLM limitations, suggest several fundamental factors that contribute to the behavioral inconsistencies observed across different contexts, tasks, and settings. We'll discuss these factors in the following section.

\section*{Challenges and Limitations of Using LLMs as Human Surrogates}

Assessing the intelligence capabilities of LLMs is challenging for several reasons, including flawed benchmarks, data contamination \citep{r54}, and inconsistent performance \citep{r86}. Nonetheless, there are several well-documented central issues and limitations of LLMs. These inconsistencies and peculiarities are often unexpected or surprising, to the point that they attract the attention of popular press.\footnote{https://www.zdnet.com/article/generative-ai-fails-in-this-very-common-ability-of-human-thought/}$^{,}$\footnote{https://futurism.com/children-destroy-ai-basic-tasks} For example, although GPT-4 has demonstrated exceptional performance on various standardized tests that assess language comprehension, coding abilities, and other skills \citep{r87}, the model has been found to struggle with some very simple, basic tasks \citep{r88,r148}. This is because LLMs are a different form of intelligence compared to human intelligence, and we do not have a good understanding yet. We discuss several essential shortcomings of LLMs next. 

\textbf{LLMs have fundamentally different objectives}: Broadly, humans have evolved over millennia with `physical grounding' in real-world stimuli \citep{r149}, which has been crucial for survival. Our instincts, or objective functions, are fundamentally different from those of LLMs, which operate with a limited, human-assigned objective function, shaped by the model's architecture and selected digitized training data. Humans and LLMs thus represent fundamentally different forms of intelligence.

\textbf{LLMs lack embodiment}: Understanding and simulating human intelligence necessitates recognizing our embodied experiences, where sensory input and physical actions are crucial. This dimension of cognition, deeply intertwined with our social and cultural contexts, is absent in LLMs. While technologically impressive, LLMs cannot capture the nuanced, embodied human intelligence aspects essential for a thorough understanding of cognition \citep{r7, r90, r124}.

\textbf{LLMs' development is running out of high-quality data:} Training Llama 3 needed over 15 trillion tokens from publicly available sources, which is seven times larger than that used for Llama 2 \citep{r158}. High-quality training datasets will become increasingly scarce, potentially creating a bottleneck for the next iteration of LLMs. Polanyi's paradox highlights that much of the tacit knowledge humans use in decision-making cannot be explicitly codified or digitized, which means that even though LLMs may excel in imitation, we lack the necessary data to enable them to learn human thought processes and understanding of the world in a complete sense \citep{r159}.

\textbf{LLMs are flawed reasoners}: They struggle with logical reasoning, a core component of human intelligence, as demonstrated by recent benchmarks \citep{r108,r109}. LLMs fail to identify accurate causal relationships despite fine-tuning efforts \citep{r110}, and they under-perform in temporal \citep{r111}, factual \citep{r112}, deductive \citep{r113}, and abstract reasoning \citep{r114}, all fundamental to human cognitive processing and decision-making. 

\textbf{LLMs' performance is erratic and lacks internal consistency:} While LLMs perform commendably on specific public datasets and benchmarks, they often falter with OOD data \citep{r108,r109}. Another example of this instability is the reverse curse \citep{r100}, where LLMs trained on statements like `A is B' often fail to comprehend `B is A' due to power law. Moreover, this inconsistency is shown in the unstable performance of LLMs on different, yet similar tasks. For instance, LLMs may excel at more challenging tasks but fail to solve simpler versions of the same task \citep{r137}. Despite being praised for their coding abilities \citep{r98}, LLMs often struggle with real-world GitHub issues \citep{r99}. Moreover, LLMs struggle with even minor modifications to ToM tasks \citep{ullman2023large}, where they typically perform well \citep{r1,r120,r121}. 

LLMs' inconsistencies are also exhibited when using in-context learning and prompt engineering techniques. CoT \citep{r144} improves LLMs' performance by deconstructing the reasoning process into intermediate steps. However, substituting these reasoning steps with meaningless tokens can also enable LLMs to tackle more complex tasks \citep{r96}. Similarly, the paper shows that randomly altering the labels of examples shown within the in-context learning does not significantly affect performance across different tasks \citep{r97}. These inconsistencies introduce substantial risks and uncertainties in evaluating and deploying LLMs. 

LLM performance is sensitive to prompting formats and \textit{how} instructions are provided. Designing effective prompts that boost LLMs' performance is fraught with challenges. Prompt brittleness refers to the idea that significant variations in LLM responses can manifest with minor changes in prompt wording \citep{r92}. Taking multiple-choice questions as an example, the number of spaces and control characters in the prompt \citep{r156}, as well as the order in which options are presented \citep{r93, r94, r95}, can greatly affect LLM performance. This variability undermines the model's reliability, especially in experimental settings where tasks are framed differently.

\textbf{The capabilities of LLMs are often overestimated due to data contamination and memorization, contributing to their inconsistency and unsatisfactory performance on new tasks:} Prior research \citep{r40,r41,r42,r43} has shown that pre-trained language models are prone to memorizing and reproducing large portions of their training data, including rare data points. This phenomenon, known as `stochastic parroting,' suggests that these models generate text by combining previously encountered patterns without genuine comprehension \citep{r92}, an idea that originated with the `Chinese Room Argument' in 1980 \citep{r155}. Consequently, models tend to excel at tasks involving familiar data but often struggle with novel tasks that fall outside their training context \citep{r46, lewis2024usingcounterfactualtasksevaluate,r109, r110}. This issue is evident in various evaluation benchmarks. One study shows that ChatGPT's memorization of the problems' answers can boost performance by over 50\% on the famous coding benchmark HumanEval \citep{r54}. Other LLMs also show inflated performance due to data memorization \citep{r52, r53, r140}. These memorization issues raise questions about whether LLM behavior patterns are due to reasoning capabilities or merely recalling specific training data points.

\textbf{LLMs are prone to hallucinations, generating content that is nonsensical or unfaithful to the questions or source materials:} This issue can arise at various stages, such as data cleaning, training, and inference \citep{r106}. It may be inevitable for current autoregressive language model architectures \citep{r107}. The occurrence of hallucinations, combined with the challenges in detecting them, poses a significant risk to the validity and reliability of research findings.

\textbf{RLHF, the essential approach to aligning LLMs with human goals, can lead to over-correction and biases in terms of simulating human surrogates:} First, RLHF is designed to make LLMs harmless and reduce toxic outputs, which limits their ability to simulate human behavior, as human thoughts and behavior are not always harmless. This alignment leads LLMs to exhibit a stronger prosocial tendency,\footnote{One of our failure mode explorations shows this.} consistent with findings that they show stronger preferences for fairness and cooperation than humans in lab experiments \citep{r103}. Second, securing high-quality feedback data for RLHF is challenging \citep{r104}. Feedback data is often homogeneous, typically sourced from a limited group of contributors, which can lead to the pigeonhole effect---oversimplification that limits diversity. While the exact impact of homogeneous feedback data on RLHF is still speculative, research has demonstrated that RLHF significantly reduces output diversity \citep{r157}. Furthermore, if the feedback data is biased, RLHF will likely amplify these biases. Recent finding indicates that ChatGPT become more politically biased after RLHF \citep{r105}.

In summary, whether LLMs can accurately simulate human-like samples remains uncertain due to many interfering factors and inherent limitations. While using LLMs to simulate human samples can be cost-effective, the potential losses from their inherent limitations could also be substantial. 

\section*{Evaluative Questions For the Researchers and Reviewers}
Based on our research findings and discussion of LLM limitations, we propose several key evaluative questions for researchers and reviewers when trying to assess and validate the use of LLMs as human surrogates or in studying human behavior.

\begin{itemize}
    \item Provide clear documentation of all experimental setups: models used (provider, version, size),\footnote{Pay close attention to using the full model name, as general names can refer to different versions over time. For instance, with OpenAI models, if you only specify `gpt-3.5-turbo' without indicating the exact version, you might be using different models, such as `gpt-3.5-turbo-0125' or `gpt-3.5-turbo-1106', depending on the time.} dates of query, interface employed (API or GUI), and prompts.
    \item Compare response distributions by running multiple iterations. Look beyond response averages and focus on the full distribution to capture nuanced differences.
    \item Known failure modes, such as hallucinations, inconsistent behavior, and reliance on memorization, must be explicitly assessed and identified.
    \item Exercise caution regarding model stability across different conditions. This includes using various models of different sizes from different providers and ideally repeating experiments over time to account for unannounced updates that may affect behavior.\footnote{Our replication results from \textit{Appendix Table \ref{replicate_without_additional},\ref{replicate_with_additional}, and \ref{replicate_brittle}} show that LLMs' performance in our task has small but statistically significant variance across time.}
    \item Test whether LLMs exhibit human-consistent task sensitivity. The model should respond to task changes in ways that align with human behavioral shifts while remaining stable when such changes would not affect human behavior. In our experiments, this was observed through variations in task descriptions, incentives, and rules.
    \item Consider prompt brittleness. Repeat the procedure with modifications to prompt text, such as changing word order or using synonyms, while keeping instructions consistent. Also, test the model's performance across multiple languages.
    \item Data leakage and memorization can impact task performance. Ask the model to generate the instructions for the experiment, as prior knowledge of these instructions may indicate exposure to previously published results. Employ different algorithms to detect data leakage, as this is an evolving area of study.
    \item Compare out-of-distribution (OOD) and in-distribution approaches to understand how models behave in unfamiliar versus familiar settings.
    \item LLMs' self-explanations provide insight into their behaviors, but these explanations may not always align with the model's actual actions.
\end{itemize}

\section*{Conclusion}
Our study uses a recent economic game to evaluate the strategic thinking of LLMs. Results reveal significant differences between humans and LLMs in reasoning depth, response distribution, and sensitivity to game framing. Notably, nearly all advanced approaches fail to make LLMs more human-like. We explore how memorization, prompt brittleness, and other failure modes affect LLM behavior. Our results reveal that LLMs' outputs are highly stochastic, shaped by the data they were trained on or exposed to, often in ways researchers cannot entirely grasp.

Our research also stresses establishing a more comprehensive, unified, robust evaluation standard in the use of LLMs in social science research. To advance research in this field, we urge scholars and reviewers to pay attention to issues such as prompt brittleness, model inconsistency, and memorization, and to conduct more robustness checks to ensure the reliability of findings. In the absence of standardized protocols, conclusions across studies lack validity and reproducibility. Researchers can easily exploit practices similar to p-hacking, such as repeatedly adjusting prompts and models to achieve desired results. In addition, LLMs may distort incentives in social science research by reducing the motivation to collect organic, high-quality data, especially in cases involving vulnerable and underrepresented communities. Even when used for preliminary studies, relying on LLM-generated synthetic data can lead to misconceptions that undermine future research efforts.

Our results could shift at any time if a model successfully memorizes the original game or the findings and data presented in our paper. Indeed, we replicated our study across different time periods, demonstrating that model outputs vary, as shown in \textit{Appendix Table \ref{replicate_without_additional},\ref{replicate_with_additional}, and \ref{replicate_brittle}}. This further emphasizes the critical need for careful documentation. Key points to remember are: 1) In-distribution or memorization performance should not be mistaken for OOD performance, which has greater implications for generalizability; 2) Performance on task \textit{A} does not guarantee similar results on task \textit{B}, regardless of how similar \textit{A} and \textit{B} may be, due to LLMs' brittleness, lack of genuine reasoning, and misalignment with human objectives; and 3) Using LLMs as a simulation testbed is fundamentally different from traditional simulations or digital twins grounded in mathematical equations and proven physical laws and curated as such. LLMs are trained on trillions of online tokens, where truth and falsehood are indistinguishable, creating a far greater disconnect from reality than in any prior scientific simulation setting.

LLMs can serve as viable tools for human behavioral research by generating novel ideas to guide hypothesis formation, as long as researchers treat them solely as tools for novelty generation, without assuming any internal or external validity in their outputs. LLMs should be viewed as tools for synthetic imagination and ideation, not as simulations of human behavior. 

Expecting to gain insights into human behavioral patterns through experiments on LLMs is like a psychologist interviewing a parrot to understand the mental state of its human owner, providing a low-resolution reflection that can easily mislead. This concern is amplified by recent studies showing that humans struggle to predict what LLMs are capable of \citep{r151}. Moreover, due to the highly capable nature of LLMs, researchers and practitioners may become trapped in an LLM echo chamber. As Ronald Coase famously remarked, ``If you torture the data long enough, it will confess to anything.” Likewise, our studies demonstrate that with extensive training and sufficient input, LLMs can be shaped to produce any desired outcome—\textit{whisper to the model long enough, and it will echo back exactly what you want to hear}. This phenomenon can also be understood as the demand effect in experimental economics, where LLMs possess a strong ability to infer the purpose of your experiment from the adjusted prompts and adapt their behavior to match the desired outcome \citep{r160}.

As we integrate LLMs into social science and business, we must recognize that LLMs, in their current state, resemble Scylla. These LLMs are human-like in appearance yet fundamentally and unpredictably different in behavior driven by their underlying multi-headed attention and auto-regressive architecture. They are \textit{Scylla Ex Machina}. Caution is essential.

\clearpage

{\fontsize{10pt}{10pt}\selectfont
\bibliography{reference}
\bibliographystyle{apalike}
\par}

\clearpage
\newpage
\pagenumbering{arabic}
\setcounter{page}{1}

\section*{Appendix}
\subsection*{Materials and Methods}
In our paper, `gpt-3.5' refers to gpt-3.5-turbo-0125, `gpt-4' refers to gpt-4-1106-preview, and `gpt-4o' refers to gpt-4o-2024-08-06. `claude3-opus' and `claude3-sonnet' correspond to claude-3-opus-20240229 and claude-3-sonnet-20240229, respectively. Llama2 and llama3 models are downloaded and deployed through Ollama. Data for the main text were collected between February and April 2024, except for the game instruction replication data, which were gathered in August 2024. All our data and codes are available at https://osf.io/m9c2a/.

The temperature setting balances coherence and diversity, with higher values producing more diverse responses. This randomness differs from human thought, which, although creative, generally maintains consistency. Therefore, a moderate temperature—ideally between 0 and 1, such as 0.5—offers a balance that enables responses that are varied yet coherent, more closely emulating human-like reasoning and decision-making. In our experiments, we set the temperature for LLM outputs to 0.5.

In our non-OOD experiment, we avoid adding system prompts or additional instructions beyond introducing the game, aiming to reduce interference from external factors. We provide only the game rules to the LLMs, requesting that they output their chosen numbers and corresponding rationales in JSON format. In the few-shot prompting setup, we include several JSON samples, each containing a choice and its rationale. Below are the instructions for the games used in our experiments and the prompts provided to the LLMs.

In the RAG experiment, we use OpenAI's Assistant API retrieval tool to embed external knowledge from the original article on the 11-20 money request game into GPT-3.5 and GPT-4. For the fine-tuning experiment, we construct a dataset including 108 human data points that match the choice distribution in the original article. We also reconstruct the reasoning for these choices based on descriptions in the article. For instance, participants choosing numbers 11 to 16 often based their choices on guesses, while those selecting numbers 17 to 20 typically provided more detailed and rational reasoning.

The hypotheses tested in the distributional analysis are formulated as follows:

\begin{align*}
H_0: E[JS_{null} - JS_{obs}] = 0; \\
H_1: E[JS_{null} - JS_{obs}] \neq 0.
\end{align*}

where \(JS_{obs}\) denotes the observed Jensen-Shannon divergence between the response distributions of the LLM and human groups, and \(JS_{null}\) represents the divergence computed under the assumption of random relabeling. This statistical test evaluates the null hypothesis that the responses from LLMs and humans have the same distribution. A low p-value would indicate significant differences in the response distributions between the LLMs and humans.

To obtain a null distribution and calculate the p-value, we first compute \(JS_{obs}\) based on the original distributions from both groups. We then perform random relabeling, ensuring the original group sizes are preserved, and calculate \(JS_{null}\) for each randomized configuration. For each iteration, we assess whether \(JS_{null} \geq JS_{obs}\). The p-value is determined by the proportion of \(JS_{null}\) values that are greater than or equal to \(JS_{obs}\), which indicates how extreme the observed divergence is compared to what we would expect under the null hypothesis. To ensure robust results, we conducted 10,000 iterations of random relabeling to approximate the permutation p-value.

\begin{center}
    \fbox{%
        \begin{minipage}{0.95\textwidth}
        \textbf{11-20 Money Request Game}
        
        \vspace{0.5em}
        \hrule
        \vspace{0.5em}
        
        \textbf{Original Version:}\\
        You and another player are playing a game in which each player requests an amount of money. 
        The amount must be (an integer) between 11 and 20 shekels. Each player will receive the amount he requests. 
        A player will receive an additional amount of 20 shekels if he asks for exactly one shekel less than the other player.
        What amount of money would you request?
        
        \vspace{0.5em}
        \hrule
        \vspace{0.5em}
        
        \textbf{Costless Version:}\\
        You and another player are playing a game in which each player chooses an integer in the range 11-20.
        A player who chooses 20 will receive 20 shekels (regardless of the other player's choice).
        A player who chooses any other number in this range will receive three shekels less than in the case where he chooses 20. 
        However, he will receive an additional amount of 20 shekels if he chooses a number that is one less than that chosen by the other player. 
        Which number would you choose?
        
        \vspace{0.5em}
        \hrule
        \vspace{0.5em}
        
        \textbf{Cycle Version:}\\
        You and another player are playing a game in which each player requests an amount of money. 
        The amount must be (an integer) between 11 and 20 shekels. Each player will receive the amount of money he requests. 
        A player will receive an additional amount of 20 shekels if (i)he asks for exactly one shekel less than the other player or (ii)he asks for 20 shekels, and the other player asks for 11 shekels. What amount of money would you request?
        \end{minipage}
    }
\end{center}

\begin{center}
\fbox{%
    \begin{minipage}{0.95\textwidth}
    \textbf{11-20 Money Request Game in Different Prompting Languages}
    \vspace{0.5em}
    \hrule
    \vspace{0.5em}
    \textbf{Spanish Version:}\\
    Estás jugando un juego con otro jugador donde cada uno solicita una cantidad específica de dinero. 
    La cantidad debe estar entre 11 y 20 shekels (números enteros). Cada jugador recibirá la cantidad exacta que solicite. Si un jugador solicita exactamente un shekel menos que el otro, recibirá adicionalmente 20 shekels.
    ¿Cuánto dinero te gustaría solicitar?
    \vspace{0.5em}
    \hrule
    \vspace{0.5em}
    \textbf{German Version:}\\
    Du und ein anderer Spieler spielen ein Spiel, bei dem jeder Spieler eine bestimmte Menge Geld anfordert. 
    Dieser Betrag muss zwischen 11 und 20 Schekel liegen (ganze Zahl). Jeder Spieler erhält den von ihm angeforderten Betrag. 
    Wenn ein Spieler genau einen Schekel weniger als der andere Spieler anfordert, erhält er zusätzlich 20 Schekel.
    Wie viel Geld möchtest du?
    \vspace{0.5em}
    \hrule
    \vspace{0.5em}
    \textbf{Chinese Version:}\\
        \begin{CJK}{UTF8}{gbsn}  
        你和另一个玩家正在玩一个游戏，每个玩家都会请求获得一定数量的钱。这个金额必须在11到20谢克尔之间(整数)。每个玩家都会收到他所请求的金额。如果一名玩家要求比另一名玩家刚好少一谢克尔，那么他将额外获得20谢克尔。你想要请求多少钱?
        \end{CJK}
    \end{minipage}
}
\end{center}

\begin{center}
\fbox{%
\begin{minipage}{0.95\textwidth}
\textbf{11-20 Money Request Game with Chain of Thought Prompting}
\vspace{0.5em}
\hrule
\vspace{0.5em}
\textbf{Zero-Shot Chain of Thought:}\\
You and another player are playing a game in which each player requests an amount of money. 
The amount must be (an integer) between 11 and 20 shekels. Each player will receive the amount he requests. 
A player will receive an additional amount of 20 shekels if he asks for exactly one shekel less than the other player.
What amount of money would you request? Let's think step by step.
\vspace{0.5em}
\hrule
\vspace{0.5em}
\textbf{Few-Shot Chain of Thought:}\\
You and another player are playing a game in which each player requests an amount of money. 
The amount must be (an integer) between 11 and 20 shekels. Each player will receive the amount he requests. 
A player will receive an additional amount of 20 shekels if he asks for exactly one shekel less than the other player.
What amount of money would you request? Let's think step by step.
Following are three examples:

\{`number': `20',` reason': `By choosing 20, I can win the most shekels that I can control regardless of other's choice.' \}

\{`number': `19', `reason': `First, I need to guess what number the other player will choose. Second, choosing the number 20 is natural 
since this is the biggest number we can choose to maximize our profit regardless of the additional 20 shekels. Second, considering 
the situation in the second step, I should choose 19 for the additional 20 shekels.' \}

\{`number': `18', `reason': `First, I need to guess what number the other player will choose. I just need to minus one based on the number, I guess, so that I can receive the additional 20 shekels. Second, the natural number a player will choose is 20 since this is the biggest number
we can choose to maximize our profit regardless of the additional 20 shekels. Third, however, I think most players will have the same belief as I do in the second step. In this case, the other player is more likely to choose 19, and I will win if I choose 18.' \}
\end{minipage}
}
\end{center}

\newpage
\subsection*{Supplementary Figures and Tables}
\begin{figure}[htbp]
\centering
\includegraphics[width=1\textwidth]{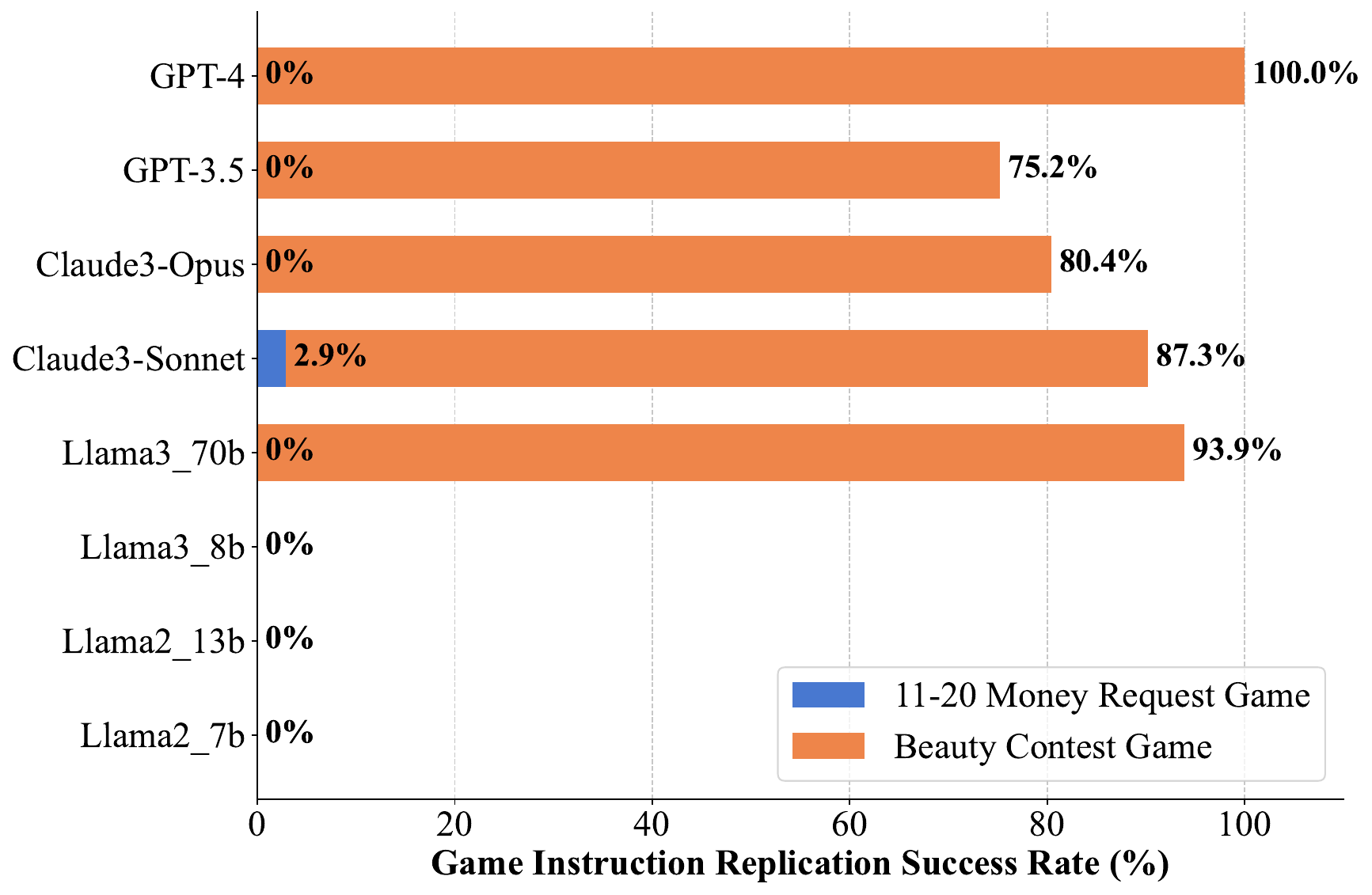}
\centering
\caption{\textbf{LLMs' Memorization on the Game instruction.} We reach out to LLMs' familiarity with the instruction of the 11-20 Money Request game and guessing game. The prompt here is in plain English: “Tell me the instructions for the 11-20 Money Request Game(2/3 Guessing Game)."}
\label{memo_instruction}
\end{figure}

\begin{figure}[htbp]
\centering
\includegraphics[width=1\textwidth]{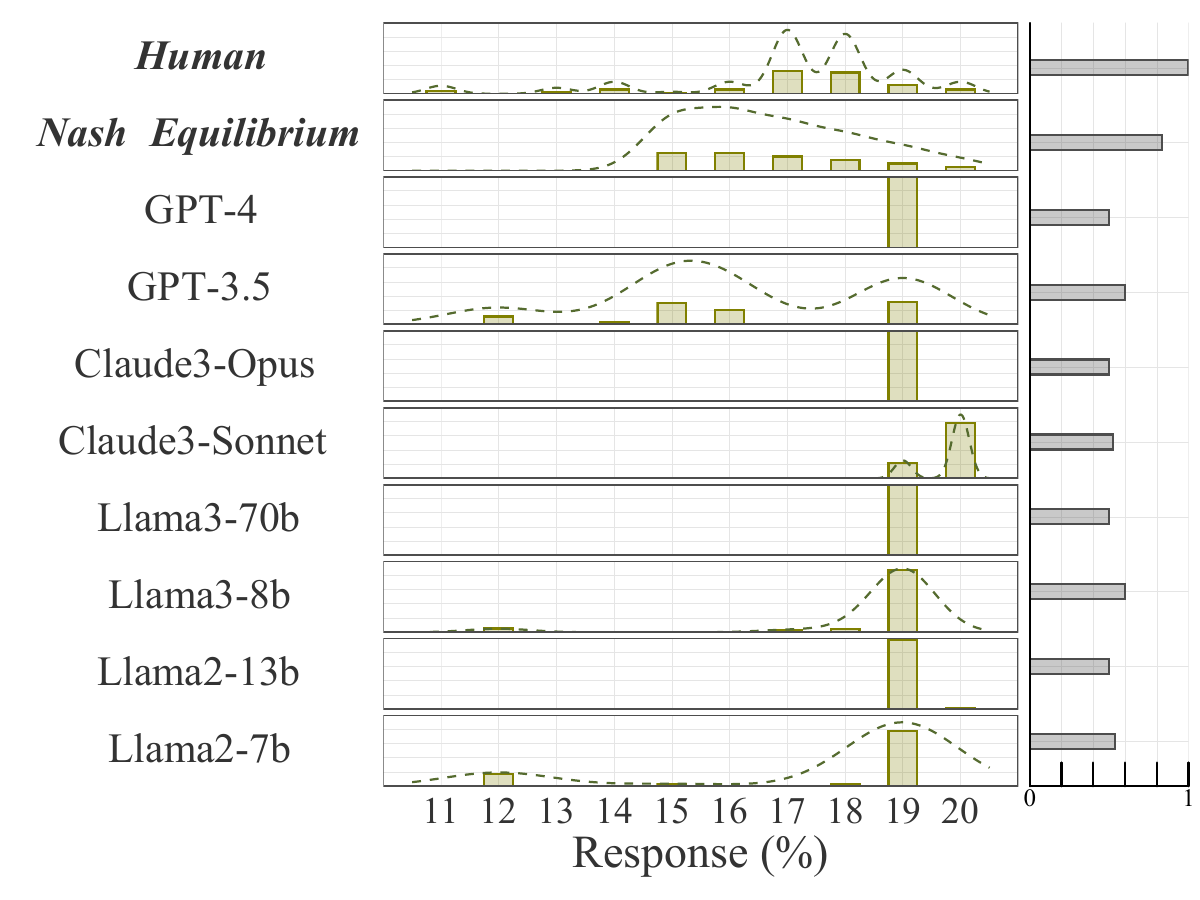}
\centering
\caption{\textbf{Advanced Prompting: Chain of Thought.} The bar chart on the right shows the similarity between the distribution of different subjects and human subjects, measured by Jensen-Shannon divergence scores. Density plots are omitted for subjects with over 98\% of the data concentrated in a single choice to avoid potential misinterpretation.}
\label{zs-cot}
\end{figure}

\begin{figure}[htbp]
\centering
\includegraphics[width=1\textwidth]{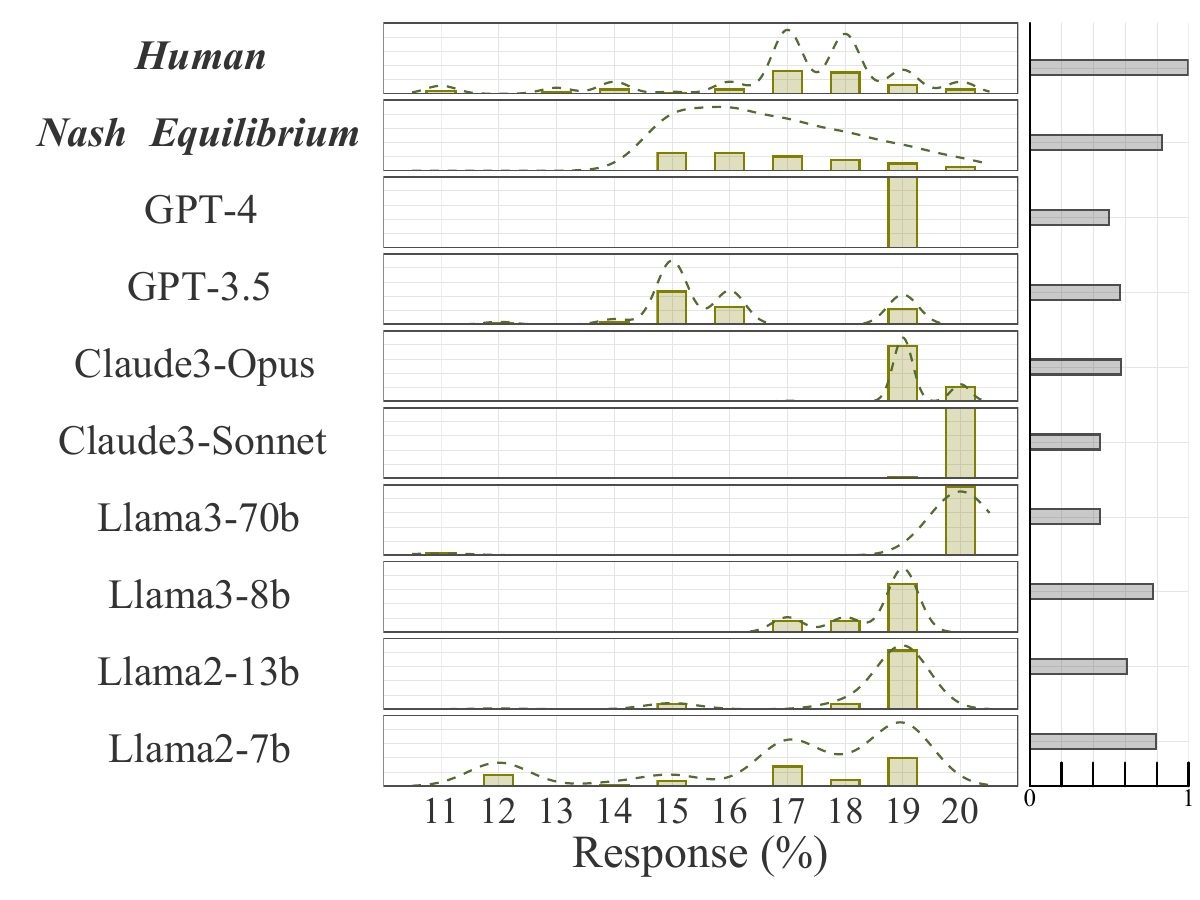}
\centering
\caption{\textbf{Advanced Prompting: Emotion Based.} The bar chart on the right shows the similarity between the distribution of different subjects and human subjects, measured by Jensen-Shannon divergence scores. Density plots are omitted for subjects with over 98\% of the data concentrated in a single choice to avoid potential misinterpretation.}
\label{zs-emotion}
\end{figure}

\newpage

\begin{figure}[!htb]
\centering
\includegraphics[width=1\textwidth]{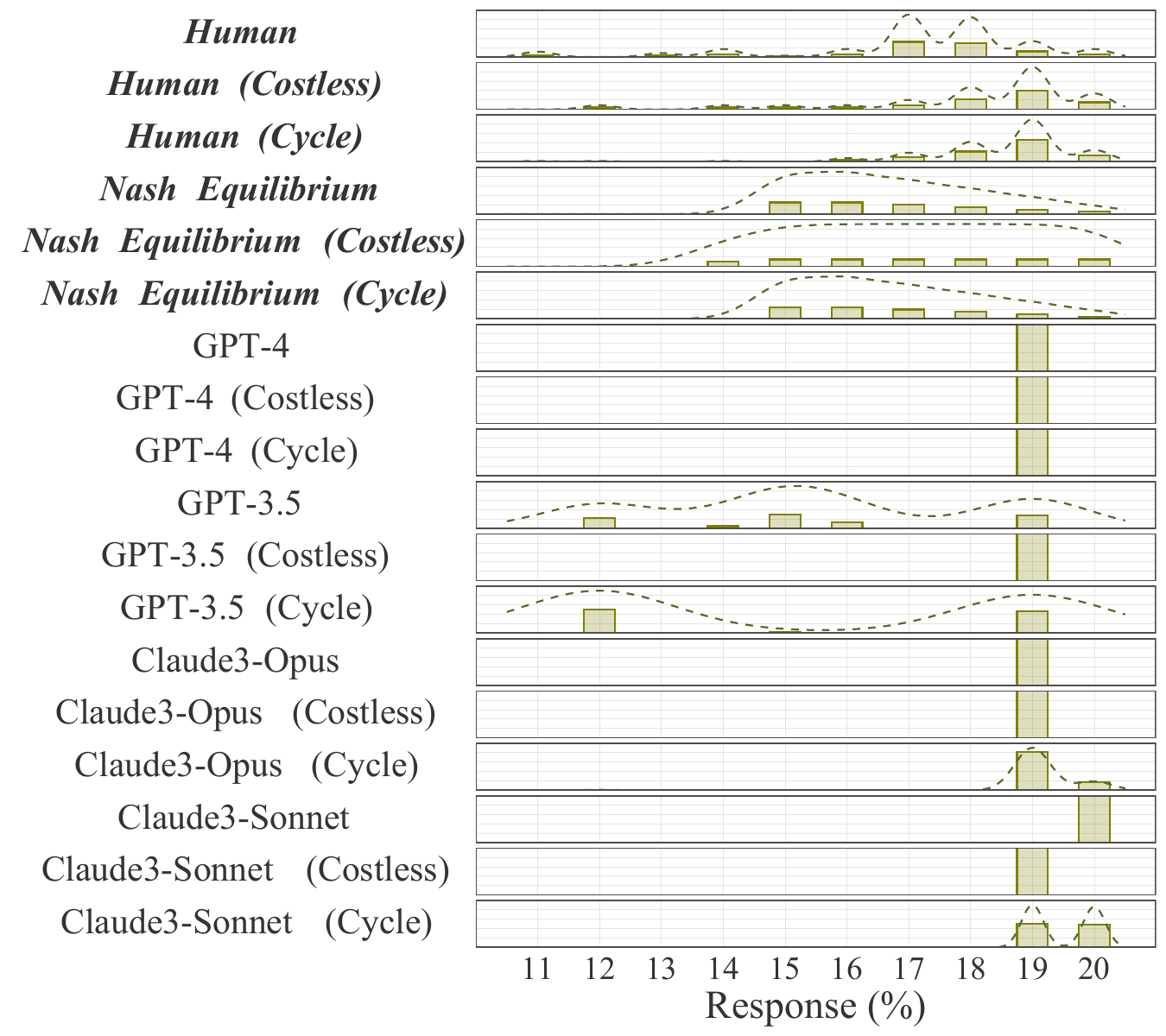}
\caption{\textbf{Humans and LLMs in Different Versions of 11-20 Money Request Game.}}
\label{diff_version}
\end{figure}



\begin{figure}[!htb]
\centering
\includegraphics[width=1\textwidth]{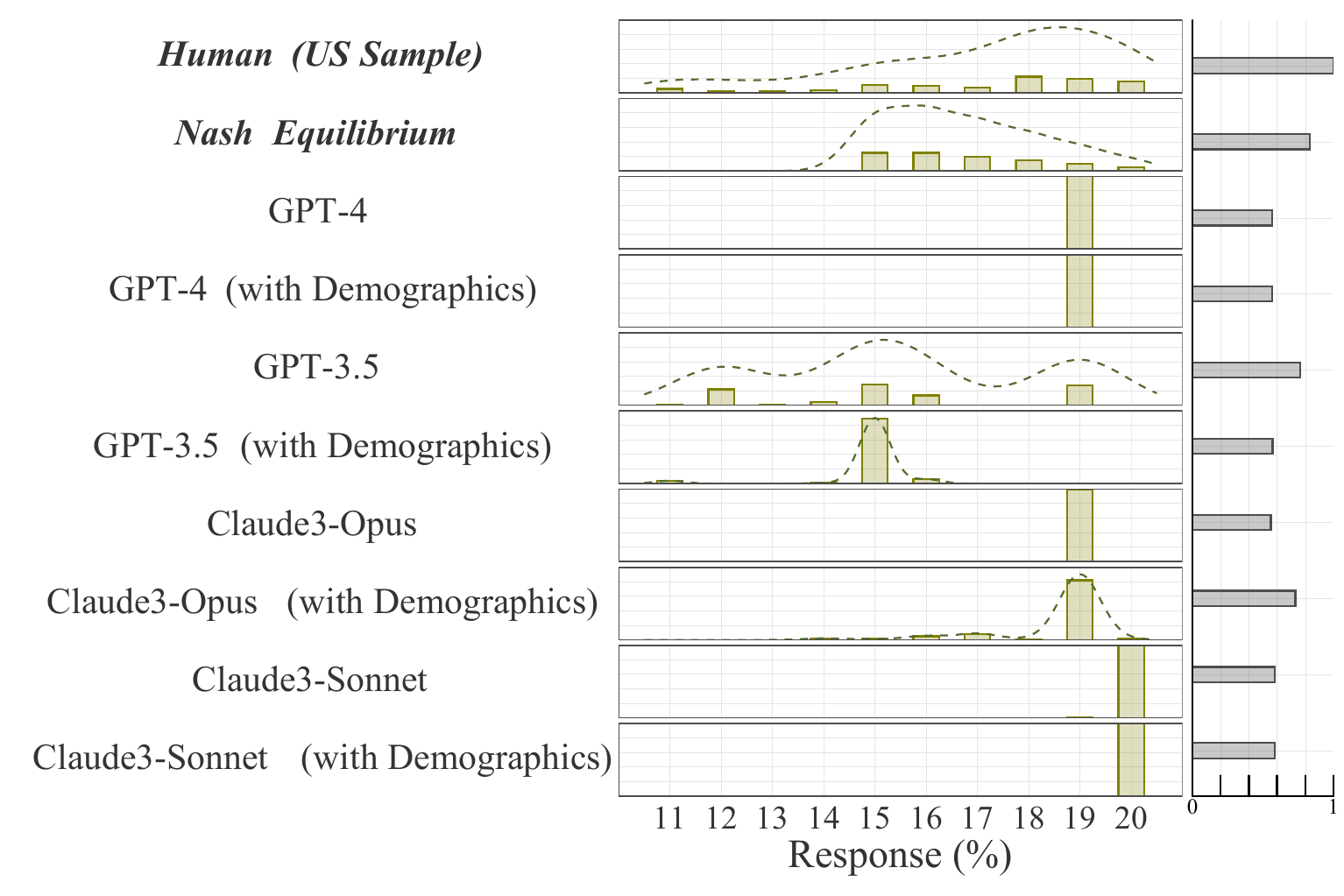}
\caption{\textbf{Additional Human Samples and LLMs' response when prompted with corresponding demographics.} The additional U.S. sample collected via Prolific exhibits similar response variances to the human samples reported in the manuscript for the 11-20 money request game. However, these U.S. samples generally demonstrate lower reasoning depth. Despite this, LLMs fail to simulate their responses, even when prompted with the corresponding demographic information.}
\label{prolific}
\end{figure}

\clearpage
\newpage
\begin{table}[ht]
\small
    \centering
    \caption{\textbf{Self-Stated Reason in 11-20 Money Request Game.} In this table, we conduct a preliminary analysis of the reasons provided by LLM subjects for their choices. For GPT-3.5 and Centaur, Only the reasons for numbers with chosen frequencies greater than 10\% are provided here. After manually categorizing these reasons, we use GPT-4-1106-preview to summarize them.}
    \label{interview}
    \begin{tabular}{|p{2cm}|c|p{10cm}|p{2cm}|}
    \hline
    \centering Subjects & Choice & \centering Reason & Failure Modes \\ \hline
    \centering
    GPT-3.5
    & 19 
    & I would request 19 shekels because if the other player requests 20 shekels, I will receive an additional 20 shekels for asking one shekel less, resulting in a total of 39 shekels. This is the maximum amount I can potentially receive in this game. \textbf{I want to maximize my potential earnings while still being eligible for the additional 20 shekels bonus.} 
    & Correct \\ \cline{3-4}
    & 
    & By requesting 19 shekels, I ensure that if the other player requests 18 shekels, I will still receive the full amount I requested without the bonus, \textbf{which is the highest amount I can request.} 
    & Misunderstand the game rules \\ \cline{2-4}
    & 16 
    & I choose to request 16 shekels because if the other player requests 17 shekels, I will receive an additional 20 shekels, making my total 36 shekels. \textbf{This is the maximum amount I can receive in this scenario.} 
    & Misunderstand the game rules \\ \cline{2-4}
    & 15 
    & I choose to request 15 shekels because if the other player requests 16 shekels, I will receive an additional 20 shekels, making my total 35 shekels. \textbf{This is the maximum amount I can receive in this game.} 
    & Misunderstand the game rules \\ \cline{2-4}
    & 12 
    & I choose to request 12 shekels because if the other player requests 13 shekels, I will receive an additional 20 shekels, making my total 32 shekels. \textbf{This is the maximum amount I can receive in this game.} 
    & Misunderstand the game rules \\ \hline
    \centering
    GPT-4 
    & 19 
    & By requesting 19 shekels, I aim to maximize my potential gain while still allowing for the possibility of earning the additional 20 shekels if the other player requests the maximum of 20 shekels, which is a common instinct. This strategy balances the risk and reward effectively, as requesting the maximum of 20 shekels would eliminate the chance of receiving the bonus, and \textbf{requesting significantly less than 19 shekels would decrease your base amount without guaranteeing the bonus.} 
    & Correct \\ \cline{3-4}
    & 
    & \textbf{By not choosing 20, I avoid the risk of the other player choosing 19 and gaining the bonus at my expense.} 
    & Misunderstand the game rules \\ \hline
    \centering
    Claude3-Opus 
    & 20 
    & Requesting 20 shekels ensures a minimum payout of 20 shekels, which is the highest guaranteed amount I can receive. Requesting 19 shekels or less risks receiving a lower payout if the other player also requests less, so \textbf{20 is the safer bet.} 
    & Fairness concern \\ \cline{3-4}
    & 
    & If the other player also requests 20 shekels, we both receive the amount we asked for, \textbf{making it a fair and likely scenario.} 
    & Fairness concern \\ \cline{3-4}
    & 
    & Should the other player request 19 shekels or less, \textbf{I would receive a total of 40 shekels (20 requested plus a 20 shekel bonus),} maximizing my potential payout. 
    & Misunderstand the game rules \\ \cline{2-4}
    & 19 
    & Requesting 19 shekels ensures the highest guaranteed minimum payout of 19 shekels, while also providing the opportunity to receive a total of 39 shekels if the other player requests 20 shekels due to the bonus for being one less. \textbf{This strategy balances the potential for the maximum bonus with a high guaranteed amount, regardless of the other player's choice.} 
    & Correct \\ \hline
    \centering
    Claude3- Sonnet 
    & 20 
    & I choose 20 shekels to guarantee the maximum base amount of money possible, which is 20 shekels, \textbf{regardless of the other player's choice.} 
    & Correct \\ \cline{3-4}
    & 
    & \textbf{If the other player requests 19 shekels, I receive an additional 20 shekels bonus, making my total 40 shekels, the highest possible payout in the game.} 
    & Misunderstand the game rules \\ \cline{3-4}
    & 
    & Requesting 20 shekels prevents the other player from gaining an advantage by requesting one shekel less than me to receive the additional bonus, thus maximizing my potential earnings \textbf{without relying on their decision.} 
    & Correct \\ \hline
    \centering
    Centaur
    & 20 
    & I chose 20 because that is the highest amount I can request.
    & Correct \\ \cline{3-4}
    & 
    & I want to earn 20 shekels; \textbf{if my opponent chooses 19, I will receive an additional reward of 20 + 20 shekels.}
    & Misunderstand the game rules \\ \cline{3-4}
    & 
    & I think my opponent will choose 11, so I choose 20.
    & Correct \\ \cline{2-4} 
    & 19 
    & The other player will probably choose 18, \textbf{so I choose 19 to get an additional 20 shekels.}
    & Misunderstand the game rulest \\  \cline{3-4}
    & 
    & I expect the other player to ask for 20, and I will win the additional 20 shekels.
    & Correct \\ \cline{2-4}
    & 18
    & I think the other player will ask for 19 and I will receive 20 shekels more.
    & Correct \\ \cline{3-4}
    & 
    & I choose 18 because \textbf{the other player is likely to choose 17 and I can win the additional bonus.}
    & Misunderstand the game rules \\ \cline{3-4}
    & 
    & The opponent will probably ask for 20, and \textbf{I will get 20 extra.}
    & Misunderstand the game rules \\ \cline{2-4} 
    & 11
    & I choose 11 because I expect my opponent will choose a higher number (probably 20) which would \textbf{give me the bonus 20 shekels.}
    & Misunderstand the game rules \\  \cline{3-4}
    & 
    & I choose 11 because it's the minimum amount I can ask for. If the other player chooses 11, we both get 11. If he chooses 12, I get 31.
    & Correct \\ \cline{2-4}
    \hline
    \end{tabular}
\end{table}

\begin{table}
\centering
\caption{\textbf{LLM Subjects' Response Variance across Different Times (Without Additional Training).} We compare the differences in means and standard deviations between the data collected in April 2024, as used in the main text of the paper, and the data collected during our replication experiment in October 2024. We also compare the distribution differences between the two data. The P-value and significance levels are obtained from 10,000 iterations of the permutation test, with the null hypothesis that any observed divergence between the two data (distributions) is due to chance. * $p<.10$, ** $p<.05$, *** $p<.01$}
\fontsize{9}{8}\selectfont
\setlength{\tabcolsep}{4pt}
\begin{tabular}{@{}l ccc c ccc c@{}}
\hline
\hline
Subjects & \multicolumn{3}{c}{Mean} & & \multicolumn{3}{c}{SD} & Distribution Difference  \\
\cline{2-4} \cline{6-8}
         & Data from& Data from& Difference & & Data from& Data from& Difference & (Measured by  \\
         & April & October & & & April & October & & JS Divergence)\\
\hline
GPT-4 & 19.00 & 19.00 & 0.00 & & 0.00 & 0.00 & 0.00 & 0.00 \\
GPT-4 (Costless) & 19.00 & 19.00 & 0.00 & & 0.00 & 0.00 & 0.00 & 0.00 \\
GPT-4 (Cycle) & 19.00 & 19.00 & 0.00 & & 0.00 & 0.00 & 0.00 & 0.00 \\
GPT-3.5 & 15.51 & 15.65 & 0.14 & & 2.57 & 1.66 & -0.91$^{***}$ & 0.07$^{***}$ \\
GPT-3.5 (Costless) & 19.00 & 19.00 & 0.00 & & 0.00 & 0.00 & 0.00 & 0.00 \\
GPT-3.5 (Cycle) & 15.34 & 15.82 & 0.48 & & 3.46 & 3.45 & -0.01 & 0.01 \\
Claude3-Opus & 19.01 & 19.06 & 0.05$^{***}$ & & 0.09 & 0.42 & 0.33$^{***}$ & 0.37$^{***}$ \\
Claude3-Opus (Costless) & 19.00 & 19.00 & 0.00 & & 0.00 & 0.00 & 0.00 & 0.00 \\
Claude3-Opus (Cycle) & 19.06 & 19.19 & 0.13 & & 0.94 & 0.39 & -0.55 & 0.01 \\
Claude3-Sonnet & 19.99 & 19.03 & -0.96$^{***}$ & & 0.09 & 0.17 & 0.08$^{**}$ & 0.60$^{***}$ \\
Claude3-Sonnet (Costless) & 19.00 & 19.00 & 0.00 & & 0.05 & 0.00 & -0.05 & 0.00 \\
Claude3-Sonnet (Cycle) & 19.46 & 19.84 & 0.38$^{***}$ & & 0.68 & 0.37 & -0.31 & 0.07$^{***}$ \\
\hline
\hline
\end{tabular}
\label{replicate_without_additional}
\end{table}

\begin{table}[htbp]
  \centering
  \caption{\textbf{LLM Subjects' Response Variance across Different Times (With Additional Training).} We compare the differences in means and standard deviations between the data collected in April 2024 and the data collected during our replication experiment in October 2024. We also compare the distribution differences between the two data. The P-value and significance levels are obtained from 10,000 iterations of the permutation test, with the null hypothesis that any observed divergence between the two data (distributions) is due to chance.  $^{*}$ $p<.10$, $^{**}$ $p<.05$, $^{***}$ $p<.01$}
\fontsize{9}{8}\selectfont
\setlength{\tabcolsep}{4pt}
\begin{tabular}{@{}l ccc c ccc c@{}}
\hline
\hline
Subjects & \multicolumn{3}{c}{Mean} & & \multicolumn{3}{c}{SD} & Distribution Difference  \\
\cline{2-4} \cline{6-8}
         & Data from& Data from& Difference & & Data from& Data from& Difference & (Measured by  \\
         & April & October & & & April & October & & JS Divergence)\\
\hline
    GPT-4-CoT (Zero-shot) & 19.00 & 19.00 & 0.00 & & 0.00 & 0.00 & 0.00 & 0.00 \\
    GPT-4-CoT (Few-shot-Large) & 12.52 & 13.46 & 0.94$^{***}$ & & 2.68 & 3.03 & 0.35$^{**}$ & 0.02$^{***}$ \\
    GPT-4-CoT (Few-shot-Small) & 14.37 & 14.00 & -0.37$^{***}$ & & 1.34 & 0.00 & -1.34$^{***}$ & 0.03$^{***}$ \\
    GPT-4o (Fine-tuned) & 17.42 & 17.90 & 0.48$^{***}$ & & 1.45 & 1.00 & -0.45$^{**}$ & 0.06$^{***}$ \\
    GPT-4-RAG & 18.49 & 18.78 & 0.29$^{***}$ & & 1.11 & 0.66 & -0.45$^{***}$ & 0.09$^{***}$ \\
    GPT-4-Emotion & 19.00 & 19.00 & 0.00 & & 0.00 & 0.00 & 0.00 & 0.00 \\
    GPT-4-Optimization & 19.00 & 19.00 & 0.00 & & 0.00 & 0.00 & 0.00 & 0.00 \\
    GPT-3.5-CoT (Zero-shot) & 16.10 & 14.76 & -1.34$^{***}$ & & 2.34 & 2.31 & -0.03 & 0.05$^{***}$ \\
    GPT-3.5-CoT (Few-shot-Large) & 17.79 & 16.91 & -0.88$^{***}$ & & 1.07 & 2.27 & 1.20$^{***}$ & 0.08$^{***}$ \\
    GPT-3.5-CoT (Few-shot-Small) & 14.00 & 14.00 & 0.00 & & 0.00 & 0.00 & 0.00 & 0.00 \\
    GPT-3.5 (Fine-tuned) & 17.41 & 17.38 & -0.03 & & 1.68 & 1.78 & 0.10 & 0.05$^{***}$ \\
    GPT-3.5-RAG & 18.54 & 17.83 & -0.71$^{***}$ & & 1.79 & 2.20 & 0.41 & 0.12$^{***}$ \\
    GPT-3.5-Emotion & 16.03 & 16.02 & -0.01 & & 1.74 & 1.68 & -0.06 & 0.00 \\
    GPT-3.5-Optimization & 16.58 & 16.22 & -0.36$^{*}$ & & 1.76 & 1.67 & -0.09 & 0.02$^{*}$ \\
    Claude3-Opus (Zero-shot) & 19.00 & 19.00 & 0.00 & & 0.00 & 0.85 & 0.85 & 0.00 \\
    Claude3-Opus (Few-shot-Large) & 17.13 & 17.04 & -0.09 & & 0.51 & 0.76 & 0.25$^{*}$ & 0.01 \\
    Claude3-Opus (Few-shot-Small) & 13.87 & 14.13 & 0.26$^{***}$ & & 0.96 & 0.61 & -0.35$^{*}$ & 0.08$^{***}$ \\
    Claude3-Opus-Emotion & 19.19 & 19.04 & -0.15$^{**}$ & & 0.51 & 0.88 & 0.37 & 0.01$^{**}$ \\
    Claude3-Opus-Optimization & 18.66 & 18.64 & -0.02 & & 1.85 & 1.66 & -0.19 & 0.02$^{**}$ \\
    Claude3-Sonnet (Zero-shot) & 19.78 & 19.88 & 0.10$^{**}$ & & 0.41 & 0.33 & -0.08$^{***}$ & 0.01$^{**}$ \\
    Claude3-Sonnet (Few-shot-Large) & 19.07 & 19.07 & 0.00 & & 0.44 & 0.33 & -0.11 & 0.00 \\
    Claude3-Sonnet (Few-shot-Small) & 16.84 & 15.56 & -1.28$^{***}$ & & 2.76 & 2.43 & -0.33$^{***}$ & 0.05$^{***}$ \\
    Claude3-Sonnet-Emotion & 19.99 & 19.95 & -0.04$^{**}$ & & 0.11 & 0.22 & 0.11 & 0.01$^{***}$ \\
    Claude3-Sonnet-Optimization & 19.84 & 19.86 & 0.02 & & 0.37 & 0.35 & -0.02 & 0.00 \\
\hline
\hline
\end{tabular}
\label{replicate_with_additional}
\end{table}

\begin{table}[htbp]
  \centering
  \caption{\textbf{LLM Subjects' Response Variance across Different Times (Under Prompt Brittleness).} We compare the differences in means and standard deviations between the data collected in April 2024 and the data collected during our replication experiment in October 2024. We also compare the distribution differences between the two data. The P-value and significance levels are obtained from 10,000 iterations of the permutation test, with the null hypothesis that any observed divergence between the two data (distributions) is due to chance. $^{*}$ $p<.10$, $^{**}$ $p<.05$, $^{***}$ $p<.01$}
\fontsize{9}{8}\selectfont
\setlength{\tabcolsep}{4pt}
\begin{tabular}{@{}l ccc c ccc c@{}}
\hline
\hline
Subjects & \multicolumn{3}{c}{Mean} & & \multicolumn{3}{c}{SD} & Distribution Difference  \\
\cline{2-4} \cline{6-8}
         & Data from& Data from& Difference & & Data from& Data from& Difference & (Measured by  \\
         & April & October & & & April & October & & JS Divergence)\\
\hline
    GPT-4 (No role/English) & 19.00 & 19.00 & 0.00 & & 0.00 & 0.00 & 0.00 & 0.00 \\
    GPT-4 (Chinese) & 19.99 & 19.99 & 0.00 & & 0.08 & 0.10 & 0.02 & 0.00 \\
    GPT-4 (Spanish) & 19.15 & 19.22 & 0.07$^{*}$ & & 0.36 & 0.42 & 0.06 & 0.004$^{*}$ \\
    GPT-4 (German) & 19.01 & 19.01 & 0.00 & & 0.12 & 0.10 & -0.02 & 0.00 \\
    GPT-4 (Rational Player) & 19.00 & 19.00 & 0.00 & & 0.00 & 0.00 & 0.00 & 0.00 \\
    GPT-4 (Human Player) & 19.00 & 19.00 & 0.00 & & 0.00 & 0.00 & 0.00 & 0.00 \\
    GPT-3.5 (No role/English) & 15.51 & 15.65 & 0.14 & & 2.57 & 1.66 & -0.91$^{***}$ & 0.07$^{***}$ \\
    GPT-3.5 (Chinese) & 17.51 & 17.86 & 0.35$^{*}$ & & 1.70 & 1.93 & 0.23$^{***}$ & 0.31$^{***}$ \\
    GPT-3.5 (Spanish) & 15.11 & 15.24 & 0.13$^{*}$ & & 0.67 & 0.95 & 0.28 & 0.01 \\
    GPT-3.5 (German) & 18.09 & 18.23 & 0.14 & & 1.68 & 1.70 & 0.02 & 0.01$^{**}$ \\
    GPT-3.5 (Rational Player) & 15.26 & 15.90 & 0.64$^{***}$ & & 0.73 & 1.78 & 1.05$^{***}$ & 0.07$^{***}$ \\
    GPT-3.5 (Human Player) & 15.27 & 15.20 & -0.07$^{*}$ & & 0.98 & 0.65 & -0.33$^{***}$ & 0.07$^{***}$ \\
\hline
\hline
\end{tabular}
\label{replicate_brittle}
\end{table}

\begin{table}[htbp]
  \centering
  \caption{\textbf{Jensen-Shannon Divergence between LLM Subjects' Distribution and Human Sample Distribution.} The p-value is calculated from 10,000 iterations of the permutation test, with the null hypothesis that any observed divergence between the LLM and human distributions is due to chance. We reject the null hypothesis in all experiment variations, indicating that LLM outputs significantly differ from human distributions, except in the fine-tuned experiments.}
  \scriptsize
  \begin{tabular}{|l|c|c||l|c|c|}
    \hline
    Subjects & Divergence & P-Value & Subjects & Divergence & P-Value \\
    \hline
    \multicolumn{6}{|l|}{\textbf{GPT-4}} \\
    \hline
    GPT-4 (English/Without Role) & 0.5025 & \textless{}0.0001 & GPT-4 (Spanish) & 0.4489 & \textless{}0.0001 \\
    GPT-4 (ZS-CoT) & 0.5024 & \textless{}0.0001 & GPT-4 (German) & 0.4853 & \textless{}0.0001 \\
    GPT-4 (FS-CoT-Large) & 0.3917 & \textless{}0.0001 & GPT-4 (Rational player) & 0.5025 & \textless{}0.0001 \\
    GPT-4 (FS-CoT-Small) & 0.4837 & \textless{}0.0001 & GPT-4 (Human player) & 0.5025 & \textless{}0.0001 \\
    GPT-4 (Emotion) & 0.5025 & \textless{}0.0001 & GPT-4o & 0.5025 & \textless{}0.0001 \\
    GPT-4 (Optimization) & 0.5025 & \textless{}0.0001 & GPT-4o (Fine-tuned) & 0.0149 & \textbf{0.3417} \\
    GPT-4 (Chinese) & 0.5645 & \textless{}0.0001 & GPT-4 (RAG) & 0.1283 & \textless{}0.0001 \\
    \hline
    \multicolumn{6}{|l|}{\textbf{GPT-3.5}} \\
    \hline
    GPT-3.5 (English/Without Role) & 0.4167 & \textless{}0.0001 & GPT-3.5 (Spanish) & 0.6160 & \textless{}0.0001 \\
    GPT-3.5 (ZS-CoT) & 0.4000 & \textless{}0.0001 & GPT-3.5 (German) & 0.4780 & \textless{}0.0001 \\
    GPT-3.5 (FS-CoT-Large) & 0.2233 & \textless{}0.0001 & GPT-3.5 (Rational player) & 0.5346 & \textless{}0.0001 \\
    GPT-3.5 (FS-CoT-Small) & 0.5779 & \textless{}0.0001 & GPT-3.5 (Human player) & 0.5283 & \textless{}0.0001 \\
    GPT-3.5 (Emotion) & 0.4332 & \textless{}0.0001 & GPT-3.5 (Fine-tuned) & 0.0302 & 0.0124 \\
    GPT-3.5 (Optimization) & 0.4284 & \textless{}0.0001 & GPT-3.5 (RAG) & 0.2040 & \textless{}0.0001 \\
    GPT-3.5 (Chinese) & 0.3818 & \textless{}0.0001 & & & \\
    \hline
    \multicolumn{6}{|l|}{\textbf{Claude3}} \\
    \hline
    Claude3-Opus & 0.4896 & \textless{}0.0001 & Claude3-Sonnet & 0.5631 & \textless{}0.0001 \\
    Claude3-Opus (ZS-COT) & 0.5024 & \textless{}0.0001 & Claude3-Sonnet (ZS-COT) & 0.4747 & \textless{}0.0001 \\
    Claude3-Opus (FS-COT-Large) & 0.2123 & \textless{}0.0001 & Claude3-Sonnet (FS-COT-Large) & 0.4060 & \textless{}0.0001 \\
    Claude3-Opus (FS-CoT-Small) & 0.4876 & \textless{}0.0001 & Claude3-Sonnet (FS-CoT-Small) & 0.3839 & \textless{}0.0001 \\
    Claude3-Opus (Emotion) & 0.4241 & \textless{}0.0001 & Claude3-Sonnet (Emotion) & 0.5581 & \textless{}0.0001 \\
    Claude3-Opus (Optimization) & 0.4088 & \textless{}0.0001 & Claude3-Sonnet (Optimization) & 0.4873 & \textless{}0.0001 \\
    \hline
    llama3-70b & 0.4233 & \textless{}0.0001 & llama3-8b & 0.3935 & \textless{}0.0001 \\
    llama3-70b (ZS-COT) & 0.5025 & \textless{}0.0001 & llama3-8b (ZS-COT) & 0.3989 & \textless{}0.0001 \\
    llama3-70b (FS-COT-Large) & 0.1609 & \textless{}0.0001 & llama3-8b (FS-COT-Large) & 0.1884 & \textless{}0.0001 \\
    llama3-70b (FS-CoT-Small) & 0.4218 & \textless{}0.0001 & llama3-8b (FS-CoT-Small) & 0.3617 & \textless{}0.0001 \\
    llama3-70b (Emotion) & 0.5557 & \textless{}0.0001 & llama3-8b (Emotion) & 0.2241 & \textless{}0.0001 \\
    llama3-70b (Optimization) & 0.5025 & \textless{}0.0001 & llama3-8b (Optimization) & 0.1850 & \textless{}0.0001 \\
    Centaur (Fine-tuned from Llama3.1) & 0.2006 & \textless{}0.0001 & & &\\
    \hline
    \multicolumn{6}{|l|}{\textbf{Llama2}} \\
    \hline
    llama2-13b & 0.5025 & \textless{}0.0001 & llama2-7b & 0.4747 & \textless{}0.0001 \\
    llama2-13b (ZS-COT) & 0.4772 & \textless{}0.0001 & llama2-7b (ZS-COT) & 0.4619 & \textless{}0.0001 \\
    llama2-13b (FS-COT-Large) & 0.2804 & \textless{}0.0001 & llama2-7b (FS-COT-Large) & 0.3276 & \textless{}0.0001 \\
    llama2-13b (FS-CoT-Small) & 0.6084 & \textless{}0.0001 & llama2-7b (FS-CoT-Small) & 0.6084 & \textless{}0.0001 \\
    llama2-13b (Emotion) & 0.3862 & \textless{}0.0001 & llama2-7b (Emotion) & 0.2074 & \textless{}0.0001 \\
    llama2-13b (Optimization) & 0.4171 & \textless{}0.0001 & llama2-7b (Optimization) & 0.4003 & \textless{}0.0001 \\
    \hline    
  \end{tabular}
  \label{js_divergence}
\end{table}

\end{document}